\documentclass{aa}
\usepackage{color}
\usepackage{graphicx}
\usepackage{orcidlink}
\usepackage{natbib}
\usepackage{txfonts}
\hypersetup{
    colorlinks=true,
    linkcolor=blue,
    citecolor=blue,
    urlcolor=blue}
    
\begin{document}
\title{Stellar Morphology of Optically Dark or Faint Galaxies at $z>3$ with JWST}

\author{
Arpita Ganguly\orcidlink{
}\inst{1,2}\thanks{E-mail: ganguly@thphys.uni-heidelberg.de}
\and
Mengyuan Xiao\orcidlink{0000-0003-1207-5344}\inst{1}\thanks{E-mail: mengyuan.xiao@unige.ch}
\and
Pascal A. Oesch\orcidlink{0000-0001-5851-6649}\inst{1,3,4}
\and
Miroslava Dessauges-Zavadsky\orcidlink{0000-0003-0348-2917}\inst{1}
\and
Andrea Weibel \orcidlink{0000-0001-8928-4465}\inst{1}
\and
Natalie Allen \orcidlink{0000-0001-9610-7950} \inst{3,4}
\and
Longji Bing\orcidlink{0000-0002-0440-7129}\inst{5}
\and
Sarah Bosman \orcidlink{0000-0001-8582-7012}\inst{2,6}
\and
Gabriel Brammer \orcidlink{0000-0003-2680-005X}\inst{3,4}
\and
David Elbaz \orcidlink{0000-0002-7631-647X} \inst{7}
\and
Emanuele Daddi \orcidlink{0000-0002-3331-9590} \inst{7}
Benjamin Magnelli \orcidlink{0000-0002-6777-6490}\inst{7}
\and
Tim B. Miller\orcidlink{0000-0001-8367-6265}\inst{8}
\and
Maxime Tarrasse\orcidlink{0009-0009-3123-4479} \inst{7}
}

\institute{
Department of Astronomy, University of Geneva, Chemin Pegasi 51, 1290 Versoix, Switzerland
\and
Institute for Theoretical Physics, Heidelberg University, Philosophenweg 12, D-69120, Heidelberg, Germany
\and
Cosmic Dawn Center (DAWN), Denmark 
\and
Niels Bohr Institute, University of Copenhagen, Jagtvej 128, DK-2200 Copenhagen N, Denmark 
\and
Astronomy Centre, University of Sussex, Falmer, Brighton, BN1 9QH UK 
\and
Max-Planck-Institut für Astronomie, Königstuhl 17, 69117 Heidelberg, Germany
\and
Université Paris-Saclay, Université Paris Cité, CEA, CNRS, AIM, 91191 Gif-sur-Yvette, France 
\and
Center for Interdisciplinary Exploration and Research in Astrophysics (CIERA), Northwestern University, 1800 Sherman Ave, Evanston, IL 60201, USA 
}

\date{Received September xxxxxx; accepted yyyyyy}

\abstract{
 The sensitivity and resolution of the James Webb Space Telescope (JWST) offer an unprecedented view of optically dark or faint galaxies (OFGs), previously missed by the Hubble Space Telescope (HST). They are candidates for massive, heavily dust-obscured star-forming galaxies (SFGs) that substantially contribute to the cosmic star formation rate density at $z > 3$. To understand what drives their high dust attenuation and how they fit into early universe galaxy evolution, we analyse the stellar morphology of 65 OFGs (selected from a parent sample of 1892 SFGs at $3<z<4$) using JWST/NIRCam F444W imaging from the PRIMER and CEERS legacy fields. We investigate correlations between dust attenuation ($A_{v}$) and key galaxy properties, including stellar mass, size, and orientation, and compare scaling relations between OFGs and typical SFGs.
We find that OFGs are $\sim$8--9 times more massive (median $\log M_\star/M_\odot=10.31^{+0.36}_{-0.42}$) and $\sim4$ times more dust attenuated (${A_{v,\mathrm{med}}=2.67 ^{+0.80}_{-0.96}}$ mag) than the parent sample. Structurally, OFGs resemble parent SFGs with median $R_e \sim 1$ kpc and median $\Sigma_{\mathrm{R_e}}\sim10^9,\mathrm{M_\odot/{kpc}^2}$. At $z>3$, OFGs may be slightly rounder on average as they prefer a higher median $q$ ($q \sim 0.60^{+0.21}_{-0.20}$) than the parent SFGs ($q \sim 0.47^{+0.22}_{-0.16}$), where, $q = b/a$, b: semi-minor axis, a: semi-major axis. While $A_v$ is strongly correlated with stellar mass, it does not show significant dependence on stellar mass-normalised effective radius ($R'_e$) and stellar mass surface density ($\Sigma'_\mathrm{{R_e}}$), Sérsic index ($n$), axis ratio ($q$), or star formation rate surface density ($\Sigma_{\mathrm{SFR}}$).  The mass-size and mass-surface density relations place OFGs as a higher-mass extension of SFGs, with no smoking-gun proof of evolutionary differences between them.
Taken together, these results suggest that OFGs are heavily dust-obscured primarily due to their high stellar masses, which facilitates significant dust production and retention, with older stellar populations likely contributing as well. Although some OFGs exhibit high $\Sigma_\mathrm{{R_e}}$ and occupy regions of the mass-size plane similar to quiescent galaxies (QGs), the overall sample is not representative of this. Their current structures resemble typical SFGs, with no concrete signs of rapid compaction. The diversity in their physical properties indicates that OFGs span a range of evolutionary states with few showing reduced star formation, while most remain actively star-forming.

}


  \maketitle

\section{Introduction}
\label{intro}
Optically dark or faint galaxies (OFGs) represent a population of massive dusty star-forming galaxies at redshifts $z$ $\geqslant$ 3. These galaxies represent a subset of dusty star-forming galaxies (DSFGs), commonly termed "HST-dark/faint" galaxies as they are often undetected or detected with low significance in even the deepest HST/WFC3 optical imaging surveys, which typically achieve a $5 \sigma$ depth of H $>$ 27 mag. This is due to their strong dust attenuation that causes a sparsity or total absence of optical/UV emission \citep[e.g.,][]{Umehata2020,Manning2022,Xiao_2023}. However, they are significantly brighter at longer wavelengths, such as Spitzer/IRAC 3.6 and 4.5 $\mu$m, JWST/NIRCam 4.44 $\mu$m, MIRI, and ALMA which enabled the identification of many such sources \citep[e.g.,][]{Wang2019,Alcalde2019, Zhou2020, Smail2021, Gomez2022a,Gomez2023,Barufet2023,Xiao_2023,2024Xiao,williams2024}.

Early research on DSFGs focused on far-infrared detections \citep[e.g.,][]{Barger1998, Hughes1998, Coppin2006, Elbaz2011, Elbaz2018A&A...616A.110E}. Understanding their stellar structure was difficult because heavy dust obscuration made them faint at optical wavelengths, and the spatial resolution of earlier instruments was insufficient. The advent of JWST has transformed this field. JWST NIRCam's subarcsecond angular resolution and exceptional sensitivity (reaching $5\sigma$ image depths of $\sim$29--30 AB mag in deep surveys), combined with coordinated large-area programs such as CEERS ($\sim$ 100 $\text{arcmin}^2$, \citealt{Bagley2023, finkelstein2023}), PRIMER ($\sim$ 200 $\text{arcmin}^2$, \citealt{Dunlop}), and COSMOS-Web ($\sim$0.6 $\text{deg}^2$, \citealt{casey2023cosmosweb}), have enabled unprecedented morphological studies of DSFGs and sub-millimeter galaxies (SMGs) \citep[e.g.,][]{Chen2022ApJ...939L...7C, Hodge2025ApJ...978..165H,Gillman2023A&A...676A..26G,mckay2025physicalpropertiesmorphologiesfaint}. Recent studies reveal that many DSFGs at $z >2$ are compact, disk-like systems with low axis ratios ($q$), and small effective radii ($R_{\mathrm{e,median}} \sim $2.7--3.2 kpc), consistent with bulge-less morphologies. A study by \cite{Gibson2024} identified a subclass of OFGs, Ultra-Flattened Objects (UFOs), in the JADES survey \citep{JADES2023arXiv231012340E}, with low $q$, Sérsic indices ($n$) close to unity and median sizes around $R_{F444W} \sim 2.7$ kpc, similar to findings on UFOs by \citet{Nelson2023}, who report median sizes of their UFO sample to be $R_\mathrm{{F444W}} \sim$ 1--2 kpc. Likewise, \citet{Gillman2024A&A...691A.299G} analysed 80 massive SMGs from the PRIMER survey ($\log M_\star/M_\odot = 11.2 \pm 0.1$), finding $R_\mathrm{{50,F444W}} = 2.7 \pm 0.2$ kpc and $n = 1.1 \pm 0.1$. \citet{mckay2025physicalpropertiesmorphologiesfaint} further confirmed these trends across $1 \leq z < 5$, reporting $R_e \sim 3.2$ kpc and $n \sim 1.1$.

 Focusing more narrowly on OFGs, a sub-class of DSFGs, \cite{Gomez2023} found 25 highly dust attenuated OFGs ($\text{A}_{\text{v} } >1$ mag) in the CEERS Survey \citep{Bagley2023, finkelstein2023} at $3 < z < 7.5$ being very compact with effective radii in F444W 30\% smaller than that of star-forming galaxies (SFGs) at the same stellar mass and redshift. These OFGs exhibit Sérsic indices $\sim$1 and axis ratios $\sim 0.5$, suggesting disk-like features. A positive correlation between dust attenuation and both size (effective radius) and stellar mass is also observed in this work. Another study by \cite{Tarrasse2025A&A...697A.181T} uses high-resolution JWST/NIRCam and HST/ACS imaging to look into massive galaxies at $3 < z < 4$. They analyse 32 RedSFGs, akin to OFGs, which are found to be compact, heavily dust-obscured systems with stellar surface densities similar to quiescent galaxies (QGs) and that likely trace a transitional phase driven by central compaction, marking an early step towards quiescence and the emergence of a "primeval bimodality" in galaxy populations. While such studies help advance our understanding of OFGs, there still exist very few studies \citep[e.g.,][]{P_rez_Gonz_lez_2023,Smail2023ApJ...958...36S,Nelson2023,Gomez2023,Gibson2024,Sun2024ApJ...961...69S,Tarrasse2025A&A...697A.181T} on the morphology of OFGs and how their stellar and dust profiles fit them into the broader framework of galaxy evolution. 
 
 Thus, in this paper, we try to delve deeper into the stellar morphology of a parent sample comprising mainly of SFGs and OFGs at 3 $<$ z $<$ 4. We address two key questions in this work.  First, how do OFGs fit into the broader framework of galaxy evolution at high redshifts? In particular, are these galaxies really structurally different from SFGs at the same epoch? Second, what physical or stellar properties drive the high dust attenuation observed in OFGs? To answer these points, we gain information on the general physical properties of OFGs in this redshift range and study the placement of OFGs within scaling relations, such as the mass-size relation and the mass-effective stellar mass density relation, which could provide hints into their potential evolutionary pathways. We then investigate the relationship between dust attenuation and galaxy properties like stellar mass, size, and axis ratio. We improve our study by incorporating a large data coverage of three legacy fields PRIMER COSMOS, PRIMER UDS and CEERS, and utilise a Bayesian light profile fitting method $\texttt{pysersic}$ to have a better estimate of uncertainties in comparison to several previous morphological analyses which have relied on $\chi^2$ fitting methods. Much care has been taken to obtain a pure OFG sample with the removal of potential quiescent galaxies (QGs) and LRDs (Little Red Dots) via various selection methods mentioned in Section \ref{Methods}.

The paper is organised into the following sections: Section \ref{Data} discusses the catalogue, SED fitting and selection methods applied to select the $3 < z < 4 $ sample used in this work. Section \ref{Methods} describes the methods used to model the light profile and measure the sizes, along with other physical properties of the selected sources. We present the results of the work in Section \ref{Results}, and finally discuss the work in Section \ref{Discussion}, followed by a conclusion in Section \ref{conclusion}.

Throughout this paper, stellar mass estimates are based on a broken power-law IMF as described in \cite{Eldridge_2017} based on \cite{kroupa}. We assume a $\Lambda$CDM cosmology with parameters $\Omega_{M}$ = 0.3, $\Omega_{\Lambda}$ = 0.7, and a Hubble constant $H_{0}$ = 70 $\text{km s}^{-1} \text{Mpc}^{-1}$.

\section{Data $\And$ Catalogue}
\label{Data}
In this section, we outline the data utilised in this work, provide details of the method of measurement of stellar properties, and describe the sample selection criteria to obtain the final sample of galaxies for our study. 
The JWST imaging data used for this work has been obtained from the CEERS (program ID: 2079, PI: S. Finkelstein; \citealt{Bagley2023,finkelstein2023}) and PRIMER surveys (program ID: 1837, PI: Dunlop, \citealt{Dunlop}) in the UDS and COSMOS fields which partially overlap with COSMOS-Web (program ID: 1727, PI: Kartaltepe $\And$ Casey, \citealt{casey2023}). We make use of the version 7.0 mosaics which are available in the DAWN JWST Archive (DJA \footnote{The DAWN JWST Archive (DJA): a repository of public JWST data, processed and reduced with grizli and msaexp, freely available to access and use: \url{https://dawn-cph.github.io/dja/}.},\cite{valentino2023}).

\subsection{Photometric Catalogue}
\label{photometric catalogue}
The primary data used in this work is taken from a photometric catalogue compiled using public JWST imaging from various Cycle 1 and 2 programs \citep{Weibel2024}. The catalogue makes use of all available data in the HST archive over the three fields in the standard ACS and WFC3/IR filters, namely ACS: F435W, F606W, F775W, F814W, and F850LP; WFC3: F105W, F125W, F140W, and F160W. Using JWST+HST data from the three survey fields, the point spread function (PSF)-matched photometric catalogues were derived using the software tool Source-Extractor \citep{SE}. The PSFs were extracted directly from the mosaics in each field. The detection image for Source-Extractor is taken as the inverse-variance weighted stack of (unconvolved) F277W, F356W, and F444W images. 

The NIRCam data in this photometric catalogue from the CEERS survey comprises 10 pointings, including the following filters: F115W, F150W, F200W, F277W, F356W, F410M, and F444W. The PRIMER survey data comprises deep JWST imaging in 10 bands: F090W, F115W, F150W, F200W, F277W, F356W, F410W, and F444W with NIRCam. The focus of this study lies in the data from NIRCam filters F150W and F444W images for color selection and F444W for morphological analyses. The total effective survey areas for each of the three fields are as follows: 82.9 $\text{arcmin}^{2}$ for CEERS with a median 5$\sigma$ depth
of 28.8 AB mag in F444W, 127.1 $\text{arcmin}^{2}$ for PRIMER-COSMOS, and 224.5 $\text{arcmin}^{2}$ for PRIMER-UDS with a median 5$\sigma$ depth of 28.4 AB mag for COSMOS and 28.2 in UDS.

\subsubsection{$\tt{EAZY}$ Redshift Measurements and $\tt{Bagpipes}$ SED Fitting}
The redshift measurements from the SED fitting code \texttt{EAZY} \citep{Brammer_2008} provided the basis for selecting galaxies at $ z >3$. The SEDs were fit over the total fluxes and errors from 14 filters: 7 in JWST/NIRCAM (F115W, F150W, F200W, F277W, F356W, F410W, F444W) and 7 in HST (F105W, F125W, F140W, F160W F435W, F606W and F814W) using the $\tt{blue\_sfhz}$ template set which comprises 13 templates created via the Flexible Stellar Population Synthesis (FSPS) code \citep{conroy2009,conroy2010}, with redshift-dependent SFHs and physical properties. Further, SED fits were performed with the Bayesian Analysis of Galaxies for Physical Inference and Parameter EStimation tool (BAGPIPES; \citealt{carnall2018}) to obtain the final redshifts and physical parameters for the selected galaxies. A delayed-$\tau$ model has been assumed for the SFH with broad uniform priors on age, the time of the beginning of star formation (0.01-5 Gyr), and on the logarithm of $\tau$ for all the galaxies in the catalogue. Stellar populations were modeled with $\tt{BPASS-v2.2}$ \citep{stanway2018}, adopting a broken power-law IMF with slopes $\alpha_{1} = -1.3$ (0.1-0.5 $M_{\odot}$) and $\alpha_{2} = -2.35$ (0.5 to 300 $M_{\odot}$) following \cite{Eldridge_2017} and \cite{kroupa}. For the dust attenuation, a Calzetti dust attenuation curve was used with a uniform prior on the dust extinction parameter $A_{v} \in (0,5)$ and on the stellar metallicity $\text{Z} \in \text{(0.1,1) Z}_{\odot}$. The final physical properties like stellar masses, redshifts, and $A_{v}$ used in this work are from BAGPIPES. For more details on the photometric catalogue and SED fitting with \texttt{EAZY} and BAGPIPES, we refer \cite {Weibel2024}

The F444W band was chosen for its effectiveness in revealing galaxies that become faint at shorter wavelengths. Targetting galaxies at $ z >3$ to $\sim$ 10, F444W, the longest wavelength band in JWST/NIRCam captures the rest-frame optical and near-infrared (NIR) flux from 493 nm (optical) to 1.11 $\mu$m (NIR), making it ideal for studying the stellar morphology of the sources. Ultimately, we limit our study to the redshift range of $3 < z < 4$ as the number of OFGs selected at $z >4$ is too small for morphological comparisons.

\begin{figure}
    \centering
    \includegraphics[width=\linewidth]{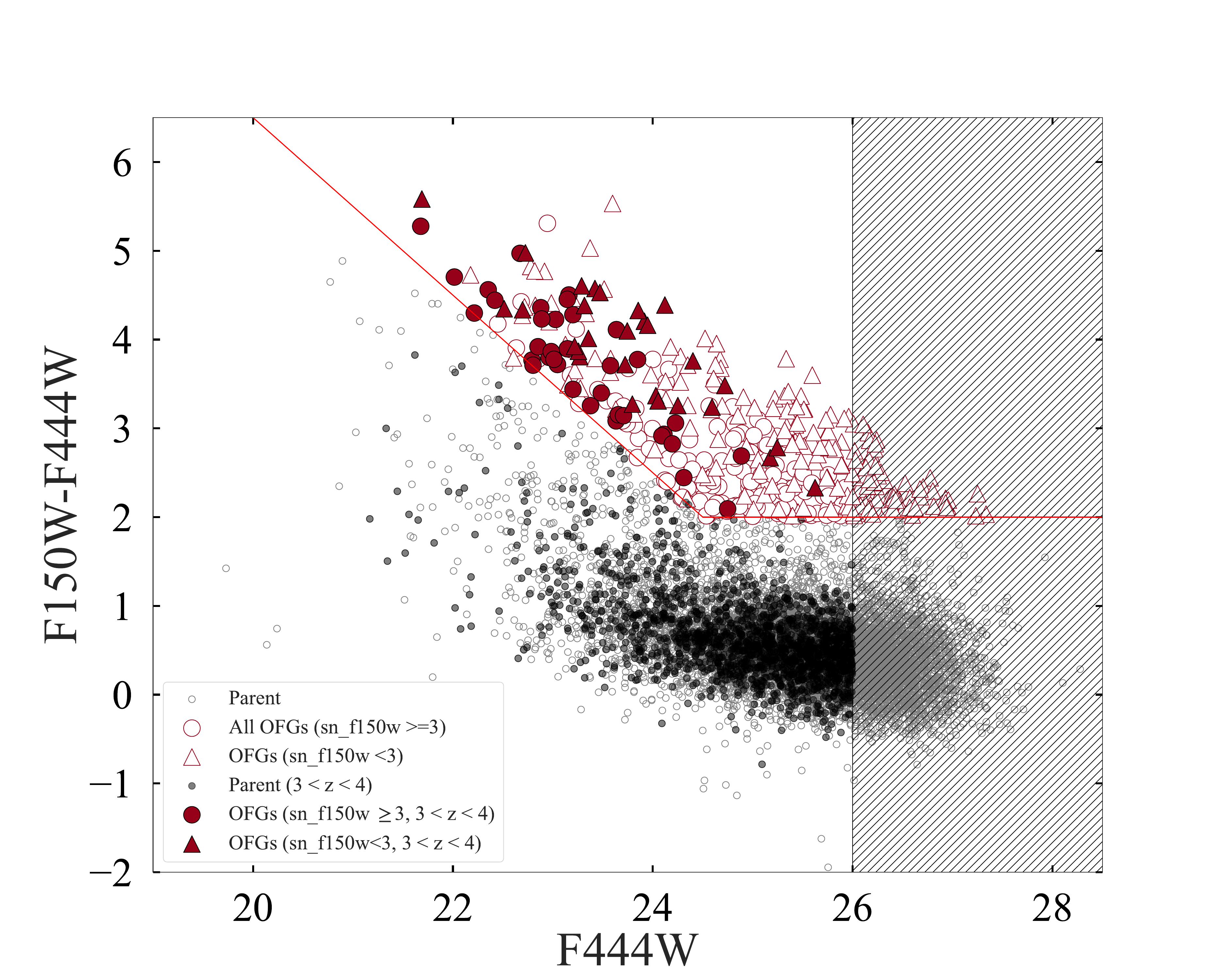}
    \caption{The sample selection from the CEERS, PRIMER-COSMOS, and PRIMER-UDS fields. Sources are identified based on their red colors in F150W-F444W versus their brightness observed in the F444W band. The red lines show the colour and magnitude cuts in F150W and F444W used to select OFGs. The grey empty circles indicate the entire parent sample of SFGs at $z > 3$. The red empty circles indicate all OFGs at $ z > 3$ which are very well detected in both F150W and F444W bands while the red empty upward triangles indicate sources with $2\sigma$ lower limits in their color. The final parent sample for $3< z < 4 $ is shown with black filled circles and the final OFG in red filled circles and filled upward triangles. Note that all OFGs are part of the parent sample. The shaded area denotes the magnitude limit of F444W $=26$ mag used for the final selection of galaxies based on results from simulation tests (see appendix \ref{pysersic_limit}). This cut allows us to recover reliable morphological measurements.}
    \label{sample}
\end{figure}
\subsection{Sample Selection}
Figure \ref{sample} provides an overview of the sample selection process. The final parent sample has been highlighted with filled black circles and the final OFG sample (a sub-sample of the parent) in filled red circles (or filled upward triangles) for the redshift range of $ 3 < z <4$. In the following subsections, we discuss in detail the steps taken to obtain these samples. 
\subsubsection{Parent sample}
 The initial sample selection relies on selecting all galaxies at $ z >3$ with a high signal-to-noise ratio (SNR) in F444W; specifically SNR(F444W) $>$ 15 to ensure that the structural properties are measured reliably. Flagged sources are removed to take into account sources of contamination such as bright saturated stars, diffraction spikes, hot pixels, and spurious detections.
 
\begin{figure}[h!]
    \centering
    \includegraphics[width=1\linewidth]{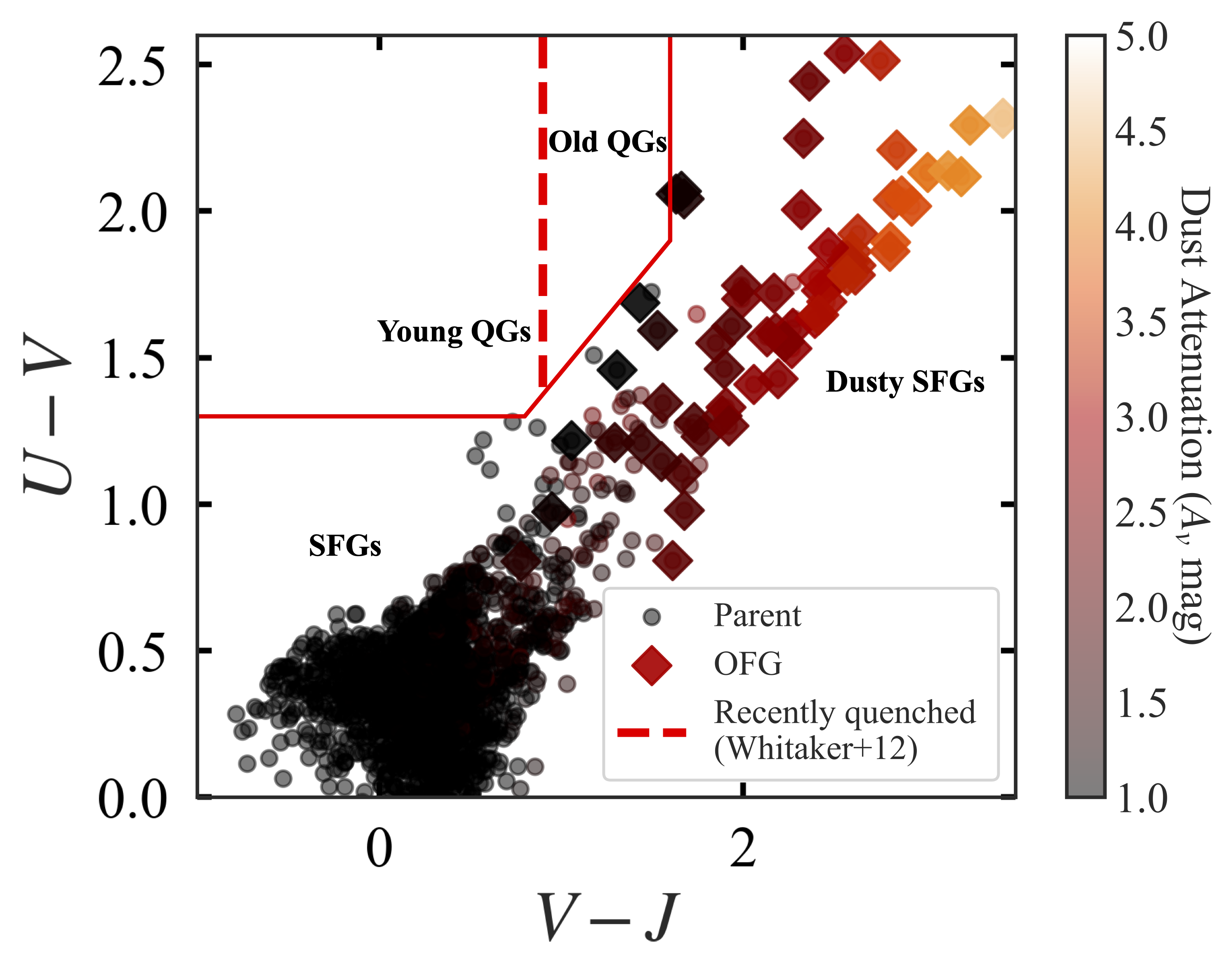}

    \caption{UVJ diagram \citep{Williams2009} for the final sample of OFGs (diamonds) and parent SFGs (dots). Sources that were found within the quiescent locus (in red) of the UVJ selection criteria, based on their rest frame colors, were removed from the final sample. The sources are colour coded with respect to their dust attenuation magnitudes ($A_{v}$ mag) obtained through methods mentioned in Section \ref{photometric catalogue}.}
    \label{fig:UVJ}
\end{figure}

 This results in 2949 sources from the PRIMER-COSMOS field, 3792 sources from the PRIMER-UDS field, and 2950 sources from the CEERS field, totalling 9691 sources. Further, based on \texttt{pysersic} simulations (see appendix \ref{pysersic_limit}), sources with flux in F444W above 26 mag are discarded due to unreliable size measurements (effective radius with >$2 \sigma$ errors). 
 To refine the sample further, we excluded sources with masks covering regions of the galaxy or that have very close neighbours which are likely to interfere with the morphological fitting. Therefore, such sources are discarded if the area within the Kron aperture contains numerous zero-valued (masked) pixels and if more than three adjacent masked pixels are detected within this aperture (see appendix \ref{kron}). This ensures that we fit relatively isolated sources which are free from contamination from the surroundings. Additionally, we performed preliminary Sérsic$+$PSF fits to exclude PSF-dominated candidates ($>$ 50\% contribution from the PSF-profile) from our final sample, as these sources could be potential AGN candidates. Since BAGPIPES SED templates do not account for AGN components in the source, this could lead to inaccurate measurements of stellar mass, redshift, and dust attenuation. PSF-dominated sources are also unsuitable for structural analyses: their observed light profiles are primarily dictated by the PSF rather than the intrinsic galaxy morphology, leading us to overestimate their compactness and result in unreliable parameter estimates. Finally, prior to fitting, we color-selected LRDs \citep[see e.g., ][]{Labbe2023b, Matthee2024, barro2024} from the parent sample, which are considered to be candidate AGNs, and we identified a notable overlap with the OFG sample (see Section \ref{OFG sample}). Since most LRDs are red, compact, and often point-like, excluding PSF-dominated sources further helps remove LRD contamination. After applying all the aforementioned selection criteria, the parent sample is reduced to 1931 sources. Furthermore, to ensure that the final samples are composed solely of non-quiescent sources, we applied a rest-frame UVJ colour cut as per the criteria discussed in \cite{Williams2009}. Among the 1931 sources, only 39 (or $\sim$2\%) were classified as QGs. Note that $\sim$ 50\% (19 out of 39) of these sources lie in the "recently quenched" or "young QGs" region \citep[e.g.,][]{Whitaker2012a,Belli2019ApJ...874...17B} of the UVJ colour diagram (see Figure \ref{fig:UVJ}), where the distinction between SFGs and QGs can be ambiguous. All QGs identified using the UVJ color criteria were therefore removed, resulting in a final parent sample of \textbf{1892} SFGs.
 
 \subsubsection{OFG Sample}
 \label{OFG sample}
 The sample of OFGs is selected based on specific magnitude and color-cuts, similar to those described in previous works \citep[e.g.,][]{Barufet2023, P_rez_Gonz_lez_2023, Gomez2023, Xiao2023a, 2024Xiao}. For our study, we have used color cuts to select OFGs that appear faint in the bluer bands. The criteria are as follows :
\begin{center}
    SNR(F444W) $>$ 15\\ \vspace{0.1 cm}
    F150W $>$ 26.5 mag \\ \vspace{0.1 cm}
    SNR(F150W) $>$ 3 \\\vspace{0.1 cm}
    F150W - F444W $>$ 2 mag\\\vspace{0.1 cm}
    If SNR(F150W) $<$ 3, 2$\sigma$ upper limit used for flux in F150W. 
\end{center}
 Using these conditions, we obtain \textbf{65} OFGs from the 1892 sources in the parent sample within $3 <z < 4$. It is notable that OFGs predominantly occupy the upper right region of the rest-frame UVJ colour space (see Figure \ref{fig:UVJ}), a location commonly associated with heavily dust-obscured SFGs \citep[e.g.,][]{Wuyts2007,Williams2009,Brammer2011ApJ...739...24B,Whitaker2015ApJ...811L..12W}. In the UVJ plane, typical QGs tend to cluster in the upper-left quadrant, while relatively unobscured SFGs appear toward the lower left. Galaxies with significant dust attenuation however, show redder U-V and V-J colours, placing them in the upper right portion of the diagram with some quite close to the locus of QGs. So the presence of OFGs in this regime prompts us to explore their dusty nature and potential role in an evolutionary pathway linking typical SFGs to quiescent systems \citep[e.g.,][]{Barro2017, Schreiber2018A&A...618A..85S, Puglisi2019, Dud2020}.

\begin{figure*}[h!]
    \centering
        \includegraphics[width=\linewidth]{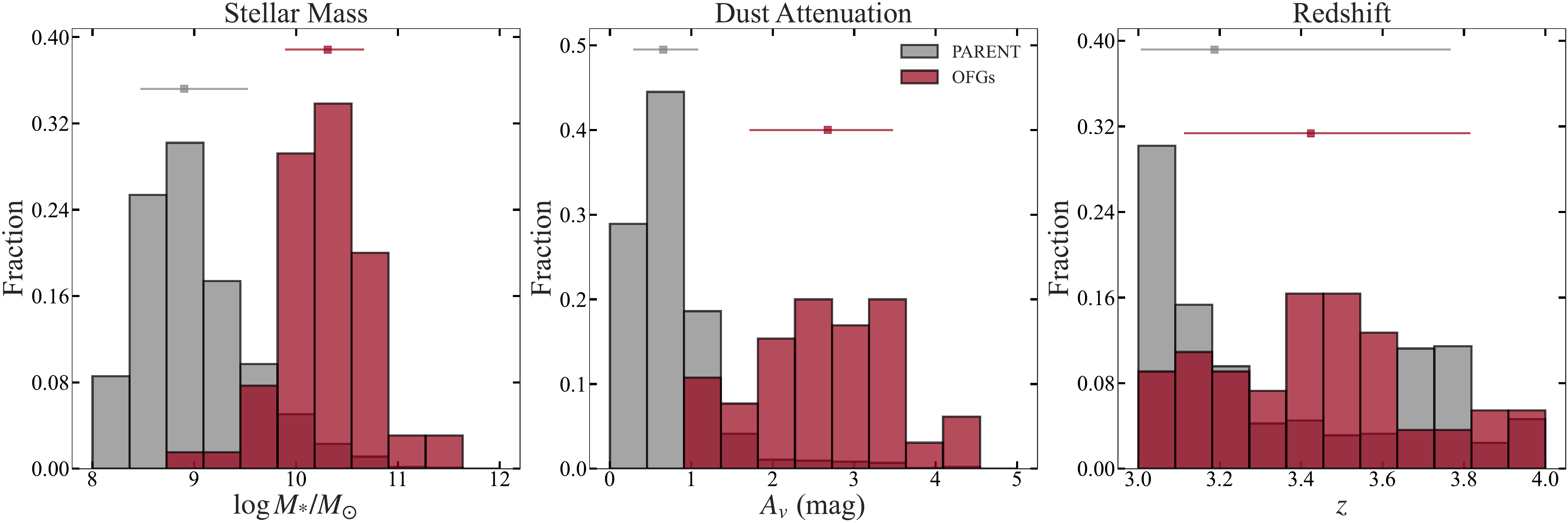}
        \label{fig:Hista}
    \vspace{0.5cm} 
        \centering
        \includegraphics[width=\linewidth]{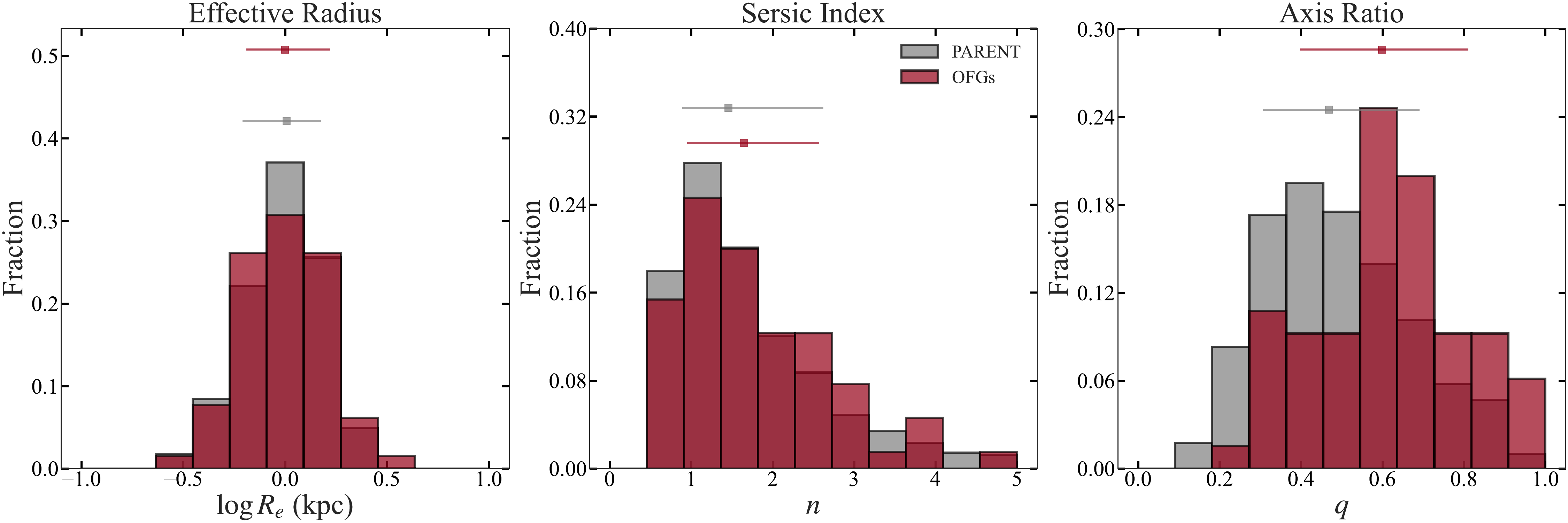}
        \label{fig:Histb}
    
    \caption{\textit{Top panel}: Histogram showing stellar mass (left), dust attenuation (middle), and redshift (right) of the parent SFG (grey) and OFG (red) samples.  \textit{Bottom panel}: Histogram showing effective radius (left), Sérsic index (middle), and axis ratio (right) of the parent SFG (grey) and OFG (red) samples. 
    For both panels, the 50th (median), 16th, and 84th percentiles of the distributions are shown in the insets above each histogram.
     The y-axis on both panels represent the fraction of the number of sources attributed to each histogram bin with respect to the total number of sources for the given population.}
    \label{fig:Combined_Hist} 
\end{figure*}
\section{Methods}
\label{Methods}
\subsection{Measurement of Galaxy Sizes and other parameters}
\label{meausrement of galaxy sizes}
We aim to study the stellar morphology of our sources by utilising the F444W JWST/NIRCam images. To do this we use $\texttt{pysersic}$ \citep{pysersic}, a recently developed Python package that uses Bayesian statistics to model Sérsic and other light profiles in astronomical images. It uses a forward modelling approach in which parametric models are fit directly to images while accounting for the effect of the PSF. In addition, we note that \texttt{GALFIT} \citep{Peng2002AJ....124..266P,Peng2010AJ....139.2097P} has been extensively used in extragalactic research to measure sizes and analyse the morphology of galaxies. During the project, through simulations, we compared the abilities of $\texttt{pysersic}$ and \texttt{GALFIT} by comparing the fluxes and PSF-to-Sérsic fractions of simulated sources representative of our real sample. Based on the fact that \texttt{pysersic} more accurately retrieves the fluxes and reproduces better-constrained PSF fractions than \texttt{GALFIT}, we conclude \texttt{pysersic} to be more efficient for our science case. \texttt{GALFIT}, nevertheless, still proves to be effective for bright, well-resolved and extended sources, a scenario not entirely reflective of our specific research needs (see appendix \ref{gal_vs_py}).

To guide the morphological fitting of light profiles in \texttt{pysersic}, we initialise priors for key parameters, such as the central coordinates (\textit{xc, yc}), flux and its uncertainty, as well as position angle ($\theta$) and its associated error. These priors are based on the values derived from Source-Extractor. For these parameters, we impose weak Gaussian priors, ensuring a well-constrained yet flexible fitting process. The central coordinates are positioned at (50, 50) pixels, corresponding to the centre of the 101 $\times$ 101-pixel cutouts, with a $\sigma$ of 1 pixel. The flux prior has a mean ($\mu$) equal to the Source-Extractor flux with a standard deviation ($\sigma$) set to twice the flux error obtained. The prior on $\theta$ is centred on the measured Source-Extractor angles with a wide Gaussian uncertainty of 60$^{\circ}$ or $\sim$1 radian. 

We set uniform priors on the rest of the physical parameters. The Sérsic index $(n)$ values are allowed to range from 0.2 to 8, the effective radii ($R_{e}$) between 1 to 20 pixels, and the axis ratios (reminder, $q = b/a$, b: semi-minor axis, a: semi-major axis) to be between 0.01 to 1.
To retrieve parameter uncertainties, we take the 50th percentile of the posterior distribution as the median, while the 16th and 84th percentiles provide the 1$\sigma$ confidence intervals for all the model parameters. Some examples of fits are presented in Appendix \ref{example} for clear visualisation.

\subsection{Presence of sub-resolution sources}
\label{sub-resolution}
Despite removing PSF-dominated sources, a significant number of Sérsic-dominated sources remain for which the median $R_e$ values fall below the resolution limit of NIRCam in F444W, i.e., the PSF's Half-Width Half Maximum (HWHM). We ensure to use this resolution limit as an upper limit to the size of such sources and consider these limits while constraining the model for the mass-size relation. This approach is supported by previous studies \citep[e.g.,][]{Vanzella2017MNRAS.467.4304V, Messa2022MNRAS.516.2420M}, which demonstrate that reliable size measurements can be obtained below the nominal resolution limit. We also tested whether the best fit scaling relations are affected by treating these aforementioned sources differently. We confirm that the relations remain unchanged regardless of whether we use the actual measured sizes of these unresolved yet Sérsic-dominated sources or adopt the PSF-HWHM resolution limit as their upper limit. Moreover, we have accounted for the uncertainties in the final measurements of data. Sources which have $>0.5$ dex and $>0.3$ dex uncertainty in stellar mass and size measurements respectively, have been discarded from the final sample. These thresholds ensure that the remaining sample retains reliable measurements. A 0.5 dex uncertainty corresponds to a factor of $\sim$ 3.2 in stellar mass, while 0.3 dex in size translates to a factor of $\sim 2$--too large of an uncertainty to be physically meaningful for our analysis, especially for sources with small sizes. 

\section{Results}
\label{Results}
\subsection{Physical Properties of OFGs}
\label{physical_prop_results}
   We begin by examining the general stellar properties of all the sources in the final sample, namely, the stellar mass ($\log M_{\star}/M_{\odot}$), dust attenuation magnitude ($A_{v}$ mag), and redshift ($z$) distribution in Figure \ref{fig:Combined_Hist} (top panel). The goal is to understand the overall distributions of these properties and identify any noticeable similarities or differences between OFGs and the parent sample. Notably, the OFGs are located at the higher mass-end of the galaxy population at $3 < z < 4$, with a median stellar mass of $\log M_{\star}/M_{\odot} = 10.31^{+0.36}_{-0.42}$. In contrast, the median stellar mass of the parent sample is $\log M_{\star}/M_{\odot} = 8.90^{+0.63}_{-0.42}$. Furthermore, when comparing $A_{v}$, OFGs exhibit a median value of $2.67 ^{+0.80}_{-0.96}$ mag, which is significantly higher, approximately $4$ times greater than that of the parent, which has a median $A_{v}$ value of $0.65^{+0.43}_{-0.37}$ mag. We check the $z$ distribution to determine if the OFG sample is representative of, or greatly different from the epoch of the larger parent sample from which it is drawn. The $z$ values for the parent sample are slightly higher at a median of $3.42^{+0.39}_{-0.31}$ than for the OFG sample ( $ z_\mathrm{median} \sim 3.18^{+0.58}_{-0.18}$), although the distribution of values for both samples is quite spread out within $3 < z < 4$. The fact that they are more massive and dust attenuated is intriguing and a motivation to study the main physical factors influencing the observed dust obscuration in OFGs. 

We then take a look at the physical model parameters which determine the light profile of the galaxies, $R_{e}$\footnote{Note that the effective radius obtained from \texttt{pysersic} fitting is measured in pixels, so we convert these values into physical units of kiloparsecs (kpc) using a cosmology-based scaling relation. Thus the physical size is calculated as:
\begin{equation}
    R_{e, \text{kpc}} = R_{e, \text{pixels}} \times \left( \frac{\text{Pixel Scale (arcsec)}}{60} \right) \times \frac{\text{kpc}}{\text{arcmin}}
\end{equation}
where we adopt the $\Lambda$CDM cosmology using \texttt{astropy.cosmology.FlatLambdaCDM} to compute the proper kpc per arcmin scale at a given redshift. }, $n$, and $q$ in Figure \ref{fig:Combined_Hist} (bottom panel). Comparing the median values of the physical properties, we find that the $R_{e}$ (parent: $1.02 ^{+1.47}_{-1.64}$ kpc, OFG: $1.00^{+1.67}_{-1.55}$ kpc) and $n$ (parent: $1.45 ^{+1.16}_{-0.57}$, OFG: $1.64^{+0.93}_{-0.70}$) distributions are similar for the two samples, while the $q$ distribution of OFGs is mildly skewed towards higher values (parent: $0.47^{+0.22}_{-0.16}$, OFG: $0.60^{+0.21}_{-0.20}$). While this difference remains within $1\sigma$, it may indicate that the OFGs on average tend to lean towards more "face-on" orientations and appear rounder in projection. The measurements indicate that OFGs may set themselves apart mainly through their higher stellar mass and $A_v$, and perhaps rounder shapes, while otherwise showing sizes and light concentration profiles comparable to the parent population.

\begin{figure*}
    \centering
    \begin{minipage}[t]{0.45\linewidth} 
        \centering
        \includegraphics[width=\linewidth]{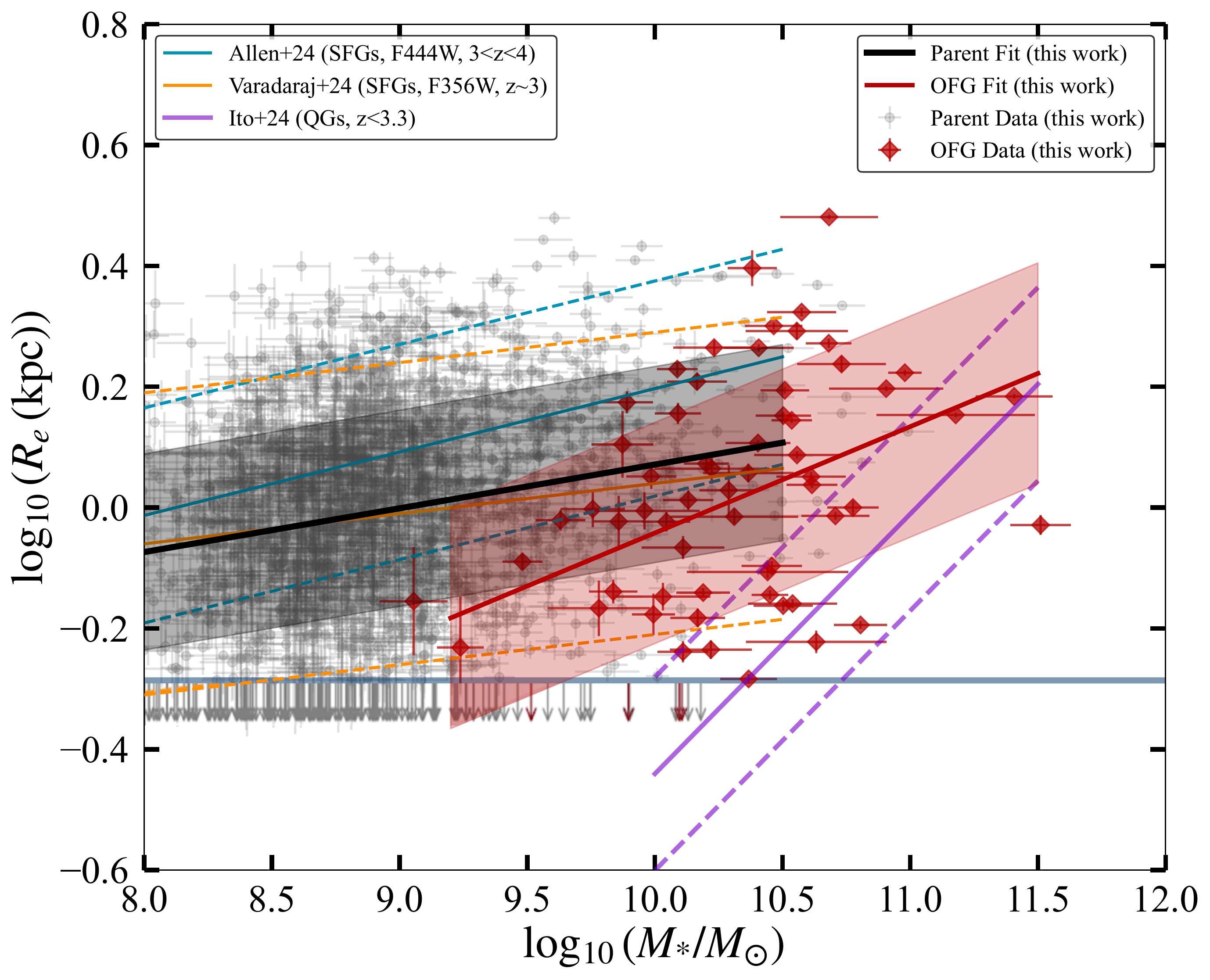}
    \end{minipage}
    \hfill
    \begin{minipage}[t]{0.5\linewidth} 
        \centering
        \includegraphics[width=\linewidth]{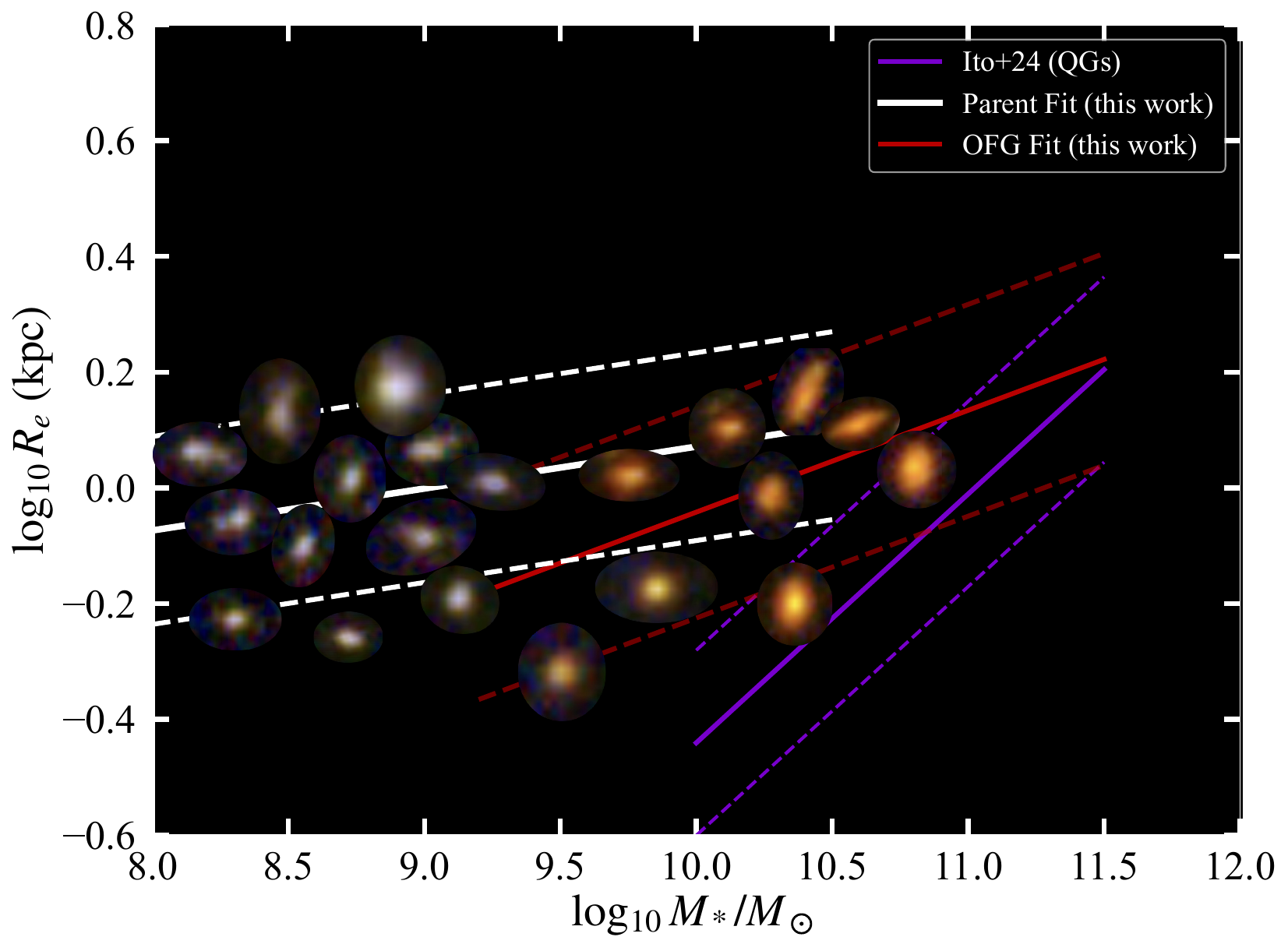}
               
    \end{minipage}

    \caption{\textit{Left panel}: Rest-frame optical mass-size relation for parent SFGs (grey points and black lines) and OFGs (red) for $3 < z < 4\, (F444W)$. The down-pointing arrows represent sources with $R_e$ smaller than the PSF HWHM or the resolution limit in F444W. These arrows indicate that the resolution limit is treated as an upper limit on their size measurements. The teal lines represent the mass-size relation at $3 < z < 4\, (F444W)$ obtained from JWST Rest-frame UV/Optical sizes \citep{allen2024galaxysizemassbuildup}, while the orange lines denote those from \cite{Varadaraj2024} at $z = 3\, (F356W)$ for reference. The purple lines indicate the mass-size relation of quiescent galaxies (QGs) at $z< 3.3$ \citep{ito2024sizestellarmass}. Dashed lines show the 1$\sigma$ scatter of the lines from the literature. The shaded light red and grey regions show the $1\sigma$ scatter of the best-fit relations of OFGs and parent SFGs, respectively. The light green line highlights the resolution limit of the NIRCam PSF in the F444W band. \textit{Right panel}: Here are sample RGB images from JWST/NIRCam (F115W, F277W, F444W) that effectively illustrate the distinctions between parent SFGs and OFGs. The mass-size relation derived for the parent sample, OFGs and QGs is also underlaid for comparison. Notably, the differences in morphology and color are striking across these categories. OFGs exhibit significantly redder colors compared to the parent sample. The parent sample displays a diverse range of morphologies, spanning from compact to more extended structures. In contrast, OFGs are characterised by predominantly round or ellipsoidal shapes. }
    \label{fig:combined-mass-size}
\end{figure*}

\subsection{The mass-size relation}
\label{mass-size section}
In this section, we try to investigate the nature of OFGs by analysing their location on the mass-size relation relative to SFGs and QGs. Our aim is to find hints about whether or not OFGs are observed at a different evolutionary phase than the parent population. Additionally, we compare the best fit values of the mass-size equation (Equation \ref{mass-size}) for the parent sample with existing studies on SFGs at comparable redshifts to assess how our result aligns with previous findings.

The mass-size relation of galaxies has been extensively studied using HST and JWST observations across low as well as high redshifts, often relying on rest-frame UV and rest-frame optical data. In the rest-frame optical wavelengths, the stellar mass-size relation for SFGs can be well described by a single power law, as shown by numerous studies \citep[e.g.,][]{Vanderwel2014,Mowla,Nedkova2021,cutler2022,Ono2024PASJ...76..219O,Varadaraj2024,Ormerod2024MNRAS.527.6110O,Morishita2024ApJ...963....9M, Martorano2024ApJ...972..134M,Miller2024arXiv241206957M,Allen2025A&A...698A..30A}.
To quantify the relation in our sample, we have used the hierarchical Bayesian linear regression tool $\tt{Linmix}$\footnote{\url{https://github.com/jmeyers314/linmix}} \citep{Kelly_2007} to fit the following mean power-law as well as an intrinsic dispersion around it for rest-optical sizes at $3 < z < 4$,
\begin{equation}
    \log R_{\text{e}} = \alpha \log \frac{M_{\star}}{M_{0}} + \log A +\sigma_\mathrm{{R_{\text{e}}}},
    \label{mass-size}
\end{equation}
where $\text{R}_{\text{e}}$ is the effective radius or the half-light radius of the galaxy, $M_{\star}$ is the stellar mass in solar masses ($M_{\odot}$), $M_{0}$ is the normalisation parameter, also measured in $M_{\odot}$; $\alpha$ is the slope, while $\log A$ denotes the intercept of the relation, which signifies the size of a galaxy at a given stellar mass $M_{0}$. Finally, $\sigma_\mathrm{R_{e}}$ is the intrinsic scatter of the relation, which reflects the expected variation in galaxy sizes. $\tt{Linmix}$ allows us to incorporate the uncertainties in both stellar mass and size measurements, providing a posterior distribution for all parameters, which is then used to derive the summary statistics--the median values and their associated confidence intervals--that characterise the underlying relation. It is worth reminding that for sources with $R_e$ below the resolution limit in the F444W band, we adopt an upper limit on their sizes corresponding to the PSF's FWHM (FWHM$/2$ at $z \sim 3.5$). This choice reflects the fact that the true sizes of such unresolved sources are unknown and likely below the resolution limit of the instrument. We incorporate this into the modeling of the mass-size relation using the $\tt{delta}$ attribute in the $\tt{Linmix}$ regression function.

Figure \ref{fig:combined-mass-size} presents the mass-size relation derived for our sources, alongside few scaling relations from the literature for comparison. The black line (with grey shading indicating the $1\sigma$ scatter) represents the scaling relation for the parent sample. The sources in the parent sample are distributed around a power-law relation with an intercept of $\log A (M_{0} = 10^{9})= -0.0011 \pm 0.0037$ and a slope of $\alpha =0.0724 \pm 0.0067$. The intrinsic scatter of the relation, $\sigma_\mathrm{{\log R_{e}}}$, is measured to be $0.1623 \pm 0.0026$. Compared with similar studies, our scaling relation appears to be slightly shallower and at a lower intercept than the F444W $3 < z < 4 $ mass-size relation for SFGs reported in \cite{Allen2025A&A...698A..30A}, which analysed unobscured SFGs in the PRIMER and CEERS fields using \texttt{GALFIT-M} \citep{Galfitm2013MNRAS.430..330H, Galfitm2013MNRAS.435..623V} within the stellar mass range of 8.5 $\leqslant \log M_{\star}/M_{\odot} \leq$ 10.5 and at 3 $< z <$ 9. By contrast, our results are quite consistent with the $ z \sim 3$ relation in the F356W band as reported by \cite{Varadaraj2024} who studied ground-based selected rest-frame UV-bright LBGs in the PRIMER field at 3 $< z <$ 5 with stellar masses $ \log M_{\star}/M_{\odot} \geqslant$ 9 using \scalebox{1}{P}\scalebox{0.8}{Y}\scalebox{1}{A}\scalebox{0.8}{UTO}\scalebox{1}{G}\scalebox{0.8}{ALAXY}, a 2-D Bayesian model fitting code. The discrepancy or similarity of our relation to aforementioned studies may stem from the distribution of the final sample used in our work, as discussed in Section \ref{mass-size-discuss}.

 The inferred parameters for the OFG mass-size relation are found to be $\alpha = 0.1760 \pm 0.0537$ and $\log A (M_{0} = 10^{9}) = -0.2179 \pm 0.0702$, with an intrinsic scatter of $0.1830 \pm 0.0156$ for the distribution of sources at 3 $< z <$ 4. To interpret these results, we must recall how SFGs and QGs behave differently on the mass-size plane. Several studies have shown that QGs exhibit a steeper mass-size relation and are more compact than SFGs of equivalent stellar mass \citep[e.g.,][]{cimatti2008, vandokkum2015,dimauro2019, Ward2024ApJ...962..176W, ito2024sizestellarmass}. This suggests that QGs undergo central compaction and subsequently grow in size primarily through dry mergers, resulting in their core dominated, compact structures \citep{Zolotov2015MNRAS.450.2327Z,Tacchella2016MNRAS.458..242T}. Figure \ref{fig:combined-mass-size} shows that the slope for the scaling relation of OFGs is steeper than the parent but shallower than the QG relation at a relatively lower redshift. Thus, the inferred mass-size relation for OFGs seems to be located at an intermediate space between that of SFGs and QGs suggesting that OFGs could be a "transitional" population between SFGs and QGs. However, a more detailed analysis of their structural properties is required to better assess the evolutionary connection among these populations. 
\\

\begin{figure}[h!]
    \centering
    \includegraphics[width=\linewidth]{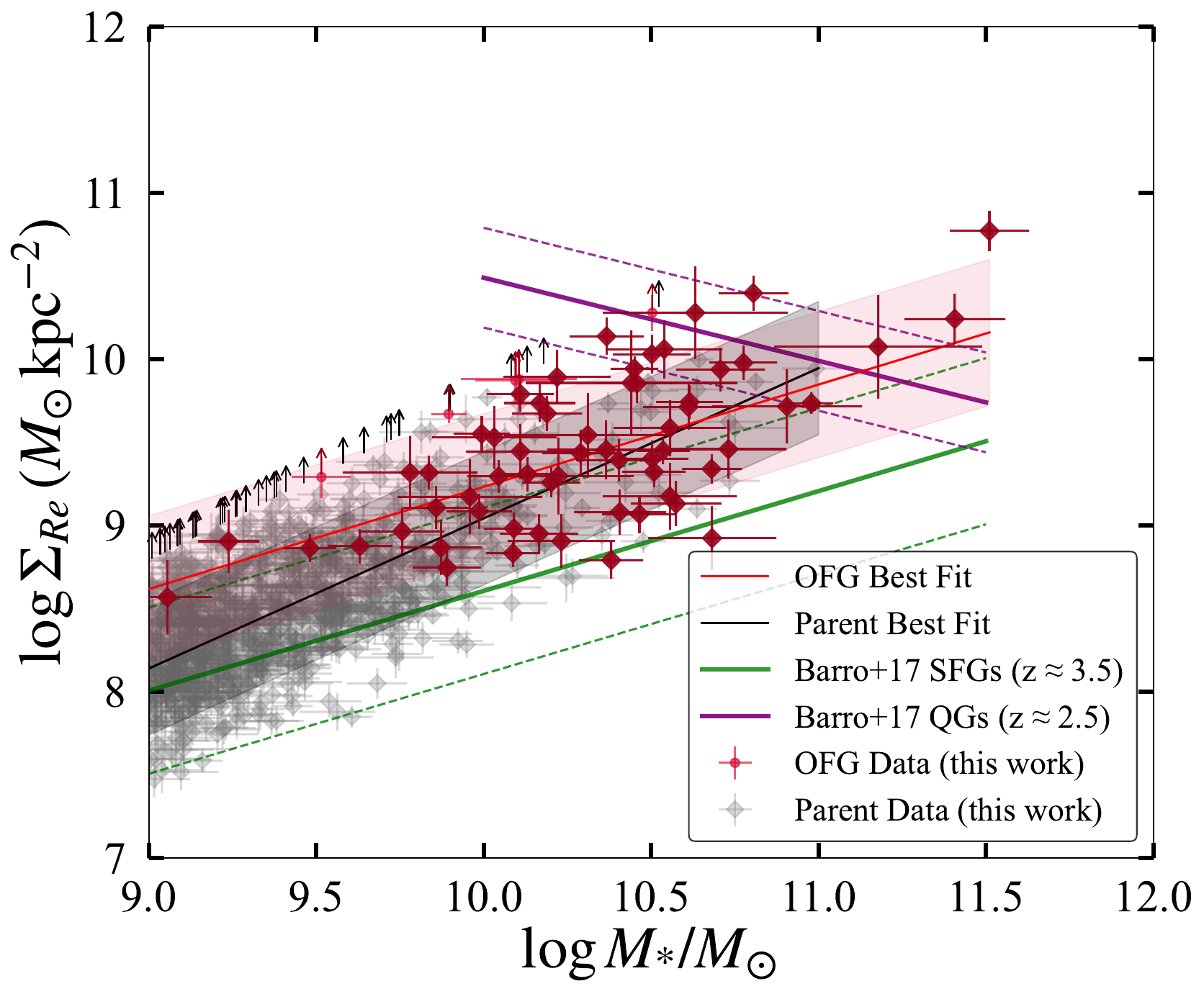}
    \caption{Redshift-binned effective stellar surface mass density $\Sigma_\mathrm{R_{e}}$ vs. stellar mass. Grey points represent SFGs, while red points represent OFGs. The upward pointing arrows represent the same sources that lie below the resolution limit in F444W as in Figure \ref{fig:combined-mass-size}. The purple solid lines indicate the $\Sigma_\mathrm{R_{e}}$-mass relation of QGs at $2 < z < 3$, with the dashed purple lines showing the intrinsic scatter for this population, as derived from \cite{Barro2017}. The green solid and dashed lines represent the redshift-extrapolated best-fit lines and intrinsic scatter for SFGs from the same study. Black lines denote the $\Sigma_\mathrm{R_{e}}$-mass relation for the parent SFG sample at the indicated redshifts (see legend). The red line represents the $\Sigma_\mathrm{R_{e}}$-mass relation for OFGs ($3 < z < 4$), with the shaded light red and grey regions showing the $1\sigma$ scatter of the best-fit relations for OFGs and parent sample, respectively.}
    \label{fig:sigma_re}
\end{figure}
\subsection{$\Sigma_{Re}$-Mass Relation}
\label{sigma_re section}

From Section \ref{physical_prop_results}, we see that OFGs tend to be more massive but have similar $R_e$ as SFGs. This suggests that they, in principle, may be more dense or "compact" than the parent sample. However, at comparable stellar masses, is this true? The results of Section \ref{mass-size section} hint that the intermediate location of OFGs in the mass-size plane could be an indication that OFGs are observed at an advanced stage of evolution in comparison to the parent sample, but can we find more proof? We try to answer these questions in this section. There exists substantial evidence that correlates the morphology of a galaxy and its stage of star formation, which in turn is indicative of its evolution \citep[e.g., ][]{Dimauro2022MNRAS.513..256D,Peterken2020MNRAS.495.3387P}. QGs often exhibit higher stellar mass densities as compared to SFGs \citep[e.g., ][]{Williams2010ApJ...713..738W, Barro2017, Tarrasse2025A&A...697A.181T,Genin2025A&A...699A.343G}. If OFGs show systematically higher effective surface mass densities ($\Sigma_\mathrm{{R_{e}}}$) than the parent sample at similar masses and occupy an intermediate position between the parent and the QG $\Sigma_\mathrm{{R_{e}}}$-mass relation, such a finding would suggest that OFGs are indeed denser systems and could be candidates for compact SFGs or serve as a transitional population, bridging the gap between SFGs and QGs. 

Figure \ref{fig:sigma_re} shows the $\Sigma_\mathrm{R_{e}}$ distribution for both the parent sample and the OFGs. $\Sigma_\mathrm{R_{e}}$ is defined as the mass enclosed within a circular area of radius equal to the galaxy's effective radius, and is given by:
\begin{equation}
\Sigma_{\text{Re}} = \frac{M_\star}{2\pi R_e^2}.
\label{simga_re_formula}
\end{equation}
The scaling relation as shown by \cite{Barro2017}) is:
\begin{equation}
\log \Sigma_{\text{Re}} = \beta \left[ \log \left( \frac{M_{\star}}{M_{\odot}} \right) - 10.5 \right] + \log B(z),
\label{sigma_re_scaling}
\end{equation}
where $\beta$ is the slope of the mass-size relation, and $\log B(z)$ is a redshift-dependent scaling constant. 

Given that OFGs are positioned at the massive end of the total sample, we focus on comparing them with SFGs within a similar stellar mass range. Therefore, we restrict this analysis to sources with $\log M_{\star}/M_\odot> 9$ to evaluate differences in compactness accurately. This leaves us with \textbf{784} sources in the parent sample, and \textbf{65} sources in the OFG sample. 

The fitting parameters derived for the $\Sigma_\mathrm{R_{e}}$-mass relation of the parent sample are $\beta = 0.904 \pm {0.070}, \ \log B(z) = 9.494 \pm{0.042}$ with an intrinsic scatter of $0.401 \pm {0.014}$. The OFGs show a shallower slope of $\beta = 0.617 \pm{0.152}, \ \log B(z) = 9.544 \pm {0.070}$ with an intrinsic scatter of $0.435 \pm {0.055}$. The OFGs are located in an area with marginally higher surface densities, with few sources occupying areas typically associated with QGs (purple slope in Figure \ref{fig:sigma_re}). Although a subset of OFGs may exhibit higher surface densities indicative of mass build-up within $R_e$, most display surface densities consistent with those of the star-forming population, suggesting more gradual stellar mass assembly processes. Moreover, as noted previously, because OFGs and SFGs exhibit identical $n$ distributions, there is no clear evidence in favour of a core-growth scenario in OFGs. So, this likely suggests that while OFGs share similar size distribution with the parent SFG population, their higher stellar masses place them at slightly elevated surface densities. 

We also note discrepancy between the $\Sigma_\mathrm{R_e}$-mass relation in our parent sample and the extrapolated relation from \cite{Barro2017} (green slope in Figure \ref{fig:sigma_re}). The relation was derived from a pre-JWST sample dominated by unobscured SFGs, and lacked the resolution now accessible with JWST. As a result, the sample may have been biased against galaxies of small sizes. Our parent sample includes a large number of small, low mass sources thanks to improved sensitivity and resolution, as well as a significant population of massive, dusty OFGs, which tend to have higher stellar masses and elevated surface densities, driving the overall $\Sigma_\mathrm{R_{e}}$-mass relation of our parent sample to a steeper slope. Additionally, the presence of numerous unresolved or marginally resolved sources in our sample contributes to a higher intercept in the fitted relation.

\subsection{Drivers of Dust Attenuation} 
\label{dust-drivers}

\begin{figure*}[h!]
    \centering

        \includegraphics[width=1\linewidth]{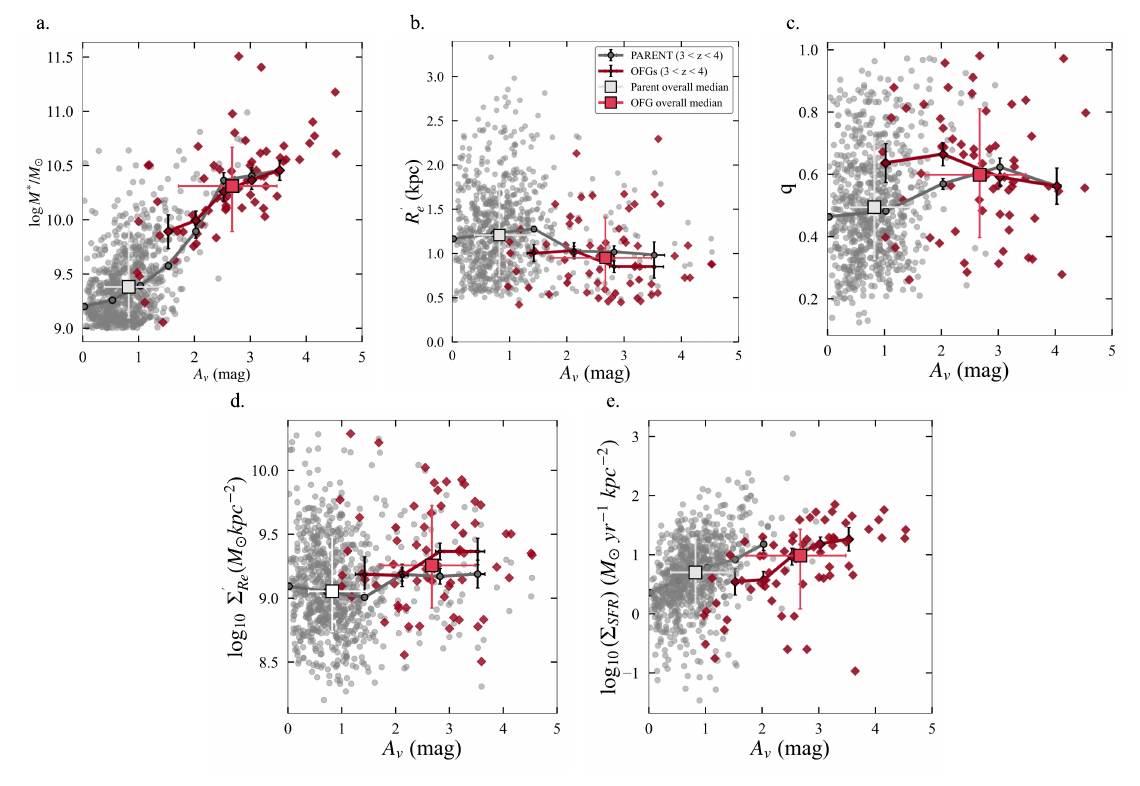}
        
    \caption{Comparison of stellar mass, stellar mass-scaled $R_{e}$ ($R'_{e}$, scaled to $\log M_{\star}/M_{\odot}$= 10), $q$, stellar mass-scaled $\Sigma_\mathrm{R_e}$ ($\Sigma'_{R_e}$, scaled to $\log M_{\star}/M_{\odot}$= 10), and star formation rate surface densities ($\Sigma_\mathrm{SFR}$) with $A_v$ of OFGs and the parent sample at $ 3 \leqslant z < 4$. Sliding medians for the two galaxy types are shown in the legend, with bold diamonds for OFGs and bold circles for parent SFGs. The error bars represent the uncertainty of the medians, and bins with fewer than ten galaxies are not displayed. For Figure \ref{fig:overall_figure_dust}b and \ref{fig:overall_figure_dust}d , the $R'_{e}$ values are $R_{e}$ values scaled to a stellar mass of $10^{10} M_{\odot}$.}
    \label{fig:overall_figure_dust}
\end{figure*}

 In this section, we aim to understand why OFGs are so heavily dust attenuated in comparison to the parent sample. Thus we try to identify the key physical properties that show correlation with $A_{v}$ to understand what impacts the observed dust attenuation among these galaxies. These physical properties can thus be addressed as the "drivers" of dust attenuation in OFGs. 
In order to have comparable mass ranges between the parent sample and OFGs, we restrict the stellar mass range of the parent SFG sample to $ \log M_{\star}/M_{\odot} \geqslant9$. In Figure \ref{fig:overall_figure_dust} we study the relationship of stellar mass, stellar mass-scaled $R_{e}$ ($R'_{e}$, scaled to $\log M_{\star}/M_{\odot}$= 10), $q$, stellar mass-scaled $\Sigma_\mathrm{R_e}$ ($\Sigma'_\mathrm{R_e}$, scaled to $\log M_{\star}/M_{\odot}$= 10), and star formation rate surface densities ($\Sigma_\mathrm{SFR}$) with $A_v$ mag at a $3 < z < 4$.

The first panel of Figure \ref{fig:overall_figure_dust} presents stellar mass as a function of $A_{v}$ mag. A strong correlation emerges between stellar mass and dust attenuation in our OFG sample, with $A_{v}$ rising sharply as stellar mass increases. This correlation is stronger as shown by a steeper $A_v - M_{\star}$ slope for OFGs than the parent sample, indicating that massive OFGs tend to be more dust-obscured. Our findings reinforce the established correlation between dust attenuation and stellar mass in SFGs, as also reported in \cite{Gomez2023}. Similar trends have been observed in previous lower redshift studies \citep[e.g.,][]{Zahid2013, Pannella2015, Alvarez2016}. Thus, our result shows that the $A_{v}- M_{\star}$ relation holds well up to $z \approx 4$.

In the second and fourth panel of Figure \ref{fig:overall_figure_dust}, we compare values of $R_{e}$ and $\Sigma_\mathrm{R_e}$, respectively, scaled to the same stellar mass of $ \log M_{\star}/M_{\odot} = 10$, with $A_{v}$ mag. This approach allows us to better understand how galaxy morphology influences dust attenuation, independent of the effects of stellar mass. The scaled values of $R'_{e}$ and $\Sigma_{R^{'}_{e}}$ are derived using the scaling factors based on the final mass-size relation we obtain for the parent SFG sample in this study (see Section \ref{mass-size section}). For completeness, we verified that using the SFG relation for SFGs and the OFG relation for OFGs produces the same results, confirming that our conclusions are not sensitive to this choice. Our results show that most sources lie at rather small normalised-sizes ($R'_\mathrm{e,med,parent} = 1.21 ^{+0.53}_{-0.46} \space \text{kpc},R'_\mathrm{e,med,OFG} = 0.95 ^{+0.46}_{-0.39} \space \text{kpc}$) with overall compact structures with large scatter ($\Sigma'_{\mathrm{R_e,med,parent}} = 9.05 ^{+0.43}_{-0.31} \space M_{\odot}\text{kpc}^{-2},\Sigma'_{R_e,med,OFG} = 9.26 ^{+0.47}_{-0.33} \space M_{\odot}\text{kpc}^{-2}$). At typical $\log M_\star$=10, the median value of $\Sigma_{R^{'}_{e}}$ of OFGs is 0.21 dex higher than the parent sample, but with significant scatter in both parent and OFG samples, the median measurements lie within the 1$\sigma$ scatter of each other. Similarly, $A_{v}$ shows a mild increase with decrease in $R'_{e}$ but not enough to imply that size might be one of the primary drivers of dust attenuation in our sample as the medians lie within the $1\sigma$ scatter.

 We do not see a significant dependence of dust attenuation values with an increase in $q$ values for either of the samples as shown in the third panel of Figure \ref{fig:overall_figure_dust}. The distribution seems almost random. This lack of correlation suggests that observed geometry of the sources do not impact dust attenuation in our sources.
We compute the star formation rate surface densities ($\Sigma_\mathrm{SFR}$) as shown in the last panel of Figure \ref{fig:overall_figure_dust}, using the formula: $ \Sigma_\mathrm{SFR}=\frac{SFR}{2 \pi R_e^{2}}$. The plot suggests a slight positive correlation between $\Sigma_\mathrm{SFR}$ and $A_{v}$ for both the parent sample and OFGs, with the distribution of both samples lying within $1\sigma$ scatter of each other. The slight rise of $\Sigma_\mathrm{SFR}$ with $A_{v}$ suggests that more dust-obscured galaxies tend to have higher $\Sigma_\mathrm{SFR}$, consistent with the idea that intense star formation is often embedded in dense, dusty regions \citep[e.g,][]{Nelson2019ApJ...870..130N}. However, since SFR strongly scales with stellar mass along the SF main sequence , part of this trend may simple reflect the underlying mass dependence rather than a direct link between dust and compact star formation. Note that the $\Sigma_\mathrm{SFR}$ refers to the dust-corrected SFR observed in these sources, which include both obscured and unobscured components. OFGs, being only visible in the NIR, their SFR is inferred from the SED fitting in NIR, incorporating correction for dust attenuation to recover the intrinsic SFR. Further interpretation of these results are discussed in Section \ref{dust_drivers_discuss}.

\section{Discussion}

\label{Discussion}
In this section, we discuss the physical implications of the main results presented in this paper: physical and stellar properties of our sample, scaling relations (Sections \ref{physical_prop_results}, \ref{mass-size section},\ref{sigma_re section}), and the relationship of stellar properties with the dust attenuation of our galaxies (Section \ref{dust-drivers}). We also compare our work with similar studies.

\subsection{Physical Properties of OFGs}
\label{physical_prop}
On comparing the stellar-based properties, namely, stellar mass and dust attenuation magnitude, we find that OFGs are both more massive and more dust attenuated than the broader SFG parent sample. This observation aligns with the color selection criteria used to select OFGs \citep[e.g.,][]{Xiao2023a}, specifically designed to target sources with moderate to high levels of dust attenuation. The morphological properties of parent SFGs and OFGs show no significant difference, except in $q$, where the distribution of OFGs is skewed towards higher values. The distribution of $q$ in both samples also provides insight into the morphologies possibly present in the galaxy samples. \cite{Kartaltepe2023ApJ...946L..15K} conducted a detailed study of galaxy morphologies at $3 < z< 9$ in the JWST CEERS survey, classifying galaxies into different structural types such as disks, spheroids, and irregulars, or combinations thereof and compared $n$, $q$, and $R_e$ across these classes. Comparing our $q$ distribution with their findings, we observe that our parent sample roughly resembles the $q$-distribution seen in disk $+$ spheroid population, suggesting a mix of extended and centrally concentrated structures. The OFGs in our sample exhibit a $q$-distribution most similar to predominantly spheroid galaxies. However, the broad $n$ distribution in both samples with a peak around 1.5 indicates the presence of disk-like or moderately concentrated light profiles in both populations. Nonetheless, given the large uncertainties, these distributions do not allow for a strict distinction between the two samples. Since the $R_e$ of both parent SFGs and OFGs share a similar distribution, the skewed $q$ in OFGs likely reflects their rounder projected morphologies relative to the parent SFG population. One plausible interpretation is that OFGs are in a transitional phase where bulge formation may be taking place, but the bulge growth argument fails since the $R_e$ and $n$ distributions for the two samples are almost identical. A higher $q$ while having same $n$ could thus imply that these galaxies are "puffed up", maintaining their light distribution while becoming more vertically extended. Such puffing up may arise if they host thicker, dynamically hotter disks, potentially driven by high internal gas fractions from continued gas accretion, by dynamical heating through minor mergers, or by other gravitational interactions that vertically thicken the disk \citep[e.g.,][]{Villalobos2008MNRAS.391.1806V,Forbes2014MNRAS.438.1552F}. Hence, from this section we can say that both the samples show significant diversity of shapes, but the OFGs stand out due to their high stellar masses, increased dust content and possibly rounder structures.
\subsection{Mass-size relation at $ 3 < z < 4$}
\label{mass-size-discuss}
\subsubsection{General mass-size relation for SFGs at $ 3 < z < 4$}
Galaxy size and morphology serve as key tracers of their evolutionary history and shed light on the processes driving stellar mass assembly and structural transformation. The location of galaxies on the mass-size plane can provide information about their evolutionary phase at the time of their observation. With this in mind, we have tried to constrain the mass-size relation for the two samples of our sources. In Section \ref{mass-size section}, we compared our findings with previous studies to assess our consistency with other works. 

The size evolution and mass-size relation for SFGs at $z > 3$ has been widely investigated as a key indicator of galaxy evolution in the early universe \citep[e.g.,][]{Ferguson2004ApJ...600L.107F,Bouwens2006ApJ...653...53B,Trujillo2006ApJ...650...18T,Mosleh2011ApJ...727....5M,Vanderwel2014}. \color{black}\cite{Shibuya2015ApJS..219...15S} compiled a large sample from HST surveys, showing a redshift-dependent size evolution where SFGs grow approximately as $R_e \propto (1+z)^{-1}$. Similarly, \cite{Mowla} extended size measurements out to $ z \approx 8$ , finding a mild evolution in the normalization (i.e., smaller sizes at higher redshifts) with limited change in the slope. With JWST, more recent works have studied the mass-size relation in the rest-optical regime up to $z \approx 9$ \citep[e.g.,][]{SunG2023MNRAS.526.2665S,Kartaltepe2023ApJ...946L..15K,Morishita2024ApJ...963....9M,Allen2025A&A...698A..30A,Varadaraj2024,Ormerod2024MNRAS.527.6110O}. 
We made a direct comparison with \cite{Allen2025A&A...698A..30A}, as both studies analyse sources in the PRIMER and CEERS fields, but differ slightly in methodology and selection criteria. \cite{Allen2025A&A...698A..30A} derive the mass-size relation for unobscured SFGs using \texttt{GALFIT-M} with a multi-wavelength fitting approach and exclude sources with $R_e$ below the PSF FWHM, considering them unresolved and their size measurements unreliable. In contrast, we retain sub-PSF sources, treating the PSF-FWHM resolution limit at F444W as an upper limit for their sizes in our fits, unless they are PSF-dominated, in which case they are removed. Many Sérsic-dominated sources remain below the empirical PSF-FWHM in the F444W band, so we do not exclude them (see Section \ref{sub-resolution}). So, the shallower slope of our best-fit relation results from the inclusion of high-mass, dust attenuated OFGs in our sample--sources that are absent in their UV-selected dataset. Meanwhile, the lower intercept of our fit is likely due to the inclusion of some unresolved sources. Thus, the difference in the final sample selection used to fit the mass size relation can explain our discrepancies with \cite{Allen2025A&A...698A..30A}. 
We then compare our result with that of \cite{Varadaraj2024}, which utilizes data from the JWST PRIMER survey to examine the mass-size scaling relations of rest-frame UV-bright Lyman-break galaxies (LBGs) at $ 3 \leqslant z < 5$ with stellar masses $\log M_{\star}/M_{\odot} >9$. Our results are highly consistent with the $ z \sim 3$ relation in the F356W band as shown in this study even when extrapolated to lower masses. This agreement is likely due to similarities in the final sample distribution. The study fits purely Sérsic and PSF$+$Sérsic profiles, to remove sources that prefer PSF$+$Sérsic fit based on their Bayesian information criterion (BIC) score to avoid contaminations from reddened AGN candidates or LRDs, which is exactly what we try to achieve by removing "PSF-dominated" sources (see Section \ref{Methods}). Thus, while our approaches slightly differ, these criteria likely lead to a comparable final sample distribution, in turn producing a highly consistent mass-size relation. 
\subsubsection{Role of OFGs in the galaxy evolution framework}
We see some differences in the estimated mass-size relation parameters for the parent SFGs and OFGs in Figure \ref{fig:combined-mass-size}. Overall, OFGs follow a steeper relation than its parent sample, a steepness that could be seen intermediate between SFGs and QGs (purple line, \citealt{ito2024sizestellarmass}). This may be an initial hint that OFGs lie in the transitional phase between SFGs and QGs, although a large scatter limits how firmly this trend can be established. To test this further, we examined the stellar mass surface densities ($\Sigma_\mathrm{R_e}$) with respect to stellar mass in Figure \ref{fig:sigma_re}. We use the redshift extrapolated fit from \cite{Barro2017} (green lines) as a reference to plot the scaling relation of $\log \Sigma_{Re}-\log M_{\star}$ for SFGs. Our parent sample of SFGs appears to deviate from this relation. These findings indicate that JWST has revealed a more compact population of SFGs at high redshift, urging the need to refine redshift extrapolations of such scaling relations. 
Within this context, OFGs occupy regions of higher $\Sigma_\mathrm{R_{e}}$, with a few sources falling along the relation associated to QGs. The slope observed in their $\Sigma_\mathrm{R_{e}}$-mass relation is not significantly different from that of the parent sample, suggesting that OFGs may not be going through rapid compaction events but gradual stellar mass growth similar to SFGs.
Taken together, these results imply that OFGs may not form a very distinct transitional class between SFGs and QGs. While a small minority show properties closer to QGs, the population overall does not show significantly increased compactness or tell-tale signs of rapid central mass build-up. The shallow slope in the $\Sigma_\mathrm{R_e}$-mass relation and their slightly elevated $\Sigma_{\mathrm{SFR}}$ suggest that many OFGs remain actively star-forming, with dust-obscured star formation occurring within their effective radii. Overall, OFGs likely represent a diverse subset of SFGs, rather than a uniform population moving toward quiescence.

\subsection{Drivers of Dust Attenuation in OFGs}
\label{dust_drivers_discuss}
We examined the drivers of dust attenuation of the galaxies to find a correlation between stellar properties like mass and size. It is true that stellar mass serves as a reliable indicator of dust attenuation, reflecting the integrated star formation history along the galaxy main sequence. As a result, more massive galaxies generally harbour higher dust and metal content. Due to their stronger gravitational potential, OFGs are better able to produce and retain dust and metals, whereas lower-mass SFGs are likely to lose these materials through galactic winds and interactions \citep[e.g.,][]{Sanchez2013,Ma2015, Genzel2015,Curti2023, Gomez2023,Looser2024}. Other possible reasons for heavy dust attenuation could be due to the presence of older stellar populations and core-collapse supernovae (CCSNe) as seen in \cite{Todini2001MNRAS.325..726T,Bocchio2016A&A...587A.157B,Schneider2024A&ARv..32....2S}. Intermediate-mass asymptotic giant branch (AGB) stars ($\log M_{\star}/M_{\odot} \sim 3-8$) can evolve into dust-producing phases within 30--100 Myr, and low-mass AGB stars ($\log M_{\star}/M_{\odot} < 3$) contribute after 150 to 500 Myr--meaning they are well in play by $z \sim $3--4 in galaxies that began forming stars at $z \geqslant6$ \citep[e.g.,][]{Gail2009ApJ...698.1136G,Ventura2012ApJ...761L..30V,Marini2023A&A...670A..97M,Tosi2023A&A...673A..41T}. Although, more studies need to be done in order understand the real distribution of older populations in these sources. Simple approaches like spatially resolved SED mapping with JWST NIRCam-NIRSpec (see \citealt{Tacchella2023MNRAS.522.6236T}), ALMA dust-continuum imaging could reveal age gradients and correlate dust clumps with older stellar populations, providing direct insight into how and where these aged stars reside within OFGs.

Some previous studies suggest that dust attenuation in each galaxy depends on the geometry of the system \citep[e.g.,][]{Giovanelli1995AJ....110.1059G, Zuckermann2021,Cochrane2024ApJ...961...37C}. Hence, among other physical properties, we investigated the effect of dust attenuation on axis ratio of the sources, and we noticed that the distribution of sources in both samples on the $A_{v}-q$ plane was close to random. Studies have shown that disk galaxies tend to show higher dust attenuation levels at higher inclinations (low $q$) due to an increased optical depth along the line of sight and thicker projected dust columns which increases the overall attenuation\citep[e.g.,][]{Wild2011MNRAS.417.1760W, patel2012ApJ...748L..27P, Gibson2024}. However, this correlation of dust attenuation and orientation remains relevant for galaxies with thin disks but loses its importance when the disks become thick \citep{Zuckermann2021}. The close to no correlation between $A_v$ and $q$ may be a hint that our sample consists of thick-disk galaxies, but further modelling of the $q$-distribution would be needed to explore this. 

Upon comparing the mass-normalised effective radii and effective surface densities, we observe that these two properties show mild trends of correlation with dust attenuation for either sample. OFGs exhibit slightly higher surface densities but remain within 1$\sigma$ of the parent distribution, while showing, approximately four times higher values of $A_{v}$ compared to parent sample (as already noted in Section \ref{physical_prop}).

Thus, in our analysis, we find a strong correlation between stellar mass and $A_{v}$, in agreement with \cite{Gomez2023}, suggesting that it is one of the primary drivers of dust attenuation. Older stellar populations and CCSne could also be likely contributors of dust seen in OFGs. Orientation shows no strong correlation with attenuation--a reason for which could be thick disk morphologies. We also see our OFGs generally have slightly lower normalized Re compared to parent sample, similar to \cite{Gomez2023} , but we note that the difference is within $1\sigma$ uncertainty.


\section{Conclusion}
\label{conclusion}
In this paper, we studied the stellar morphology of SFGs and OFGs within the redshift range of $3 < z < 4$. We studied in detail the mass-size relation and stellar mass surface density-mass relation for the two populations. We then examined the relationship of dust attenuation magnitude with global stellar properties to identify correlations among them and find differences between the two samples. The following are the main takeaways from this work.

\begin{enumerate}
    \item The selection criteria used in this work effectively selected high redshift SFGs and OFGs with low contamination from QGs (2\%). After removal of QGs and PSF-dominated sources, we analyse 65 OFGs from a parent sample of 1892 SFGs at $3 <z <4$.
    \item OFGs are more massive and $\sim$4 times more dust attenuated than SFGs with stellar masses $> 9 \times 10^{10}$ ($\log M_\mathrm{\star,med} \sim 10.31$) and dust attenuation magnitudes of $3< z < 4$ ${A}_{v,\mathrm{med}} \sim $ 2.67. For both color-selected populations, the sizes appear small ${R}_{\mathrm{e, med}} \sim 1 \mathrm{kpc}$ and portray disk like structure with Sérsic indices around $n_{\mathrm{ med}} \sim 1.5 $.
    
    \item For stellar masses $> 9 \times 10^{10} M_{\odot}$, stellar mass remains the main proxy for dust attenuation, among all the physical properties studied. While weak correlations are found with effective radii (scaled to the same stellar mass of $\log M_{\star}/M_{\odot}=10$), axis ratios, stellar mass surface densities (similarly scaled), and star formation rate surface densities, all quantities remain consistent within $1\sigma$ uncertainty of each other.

    \item Older stellar populations may be present in OFGs which may contribute to elevated dust content in these galaxies but their distribution within the galaxy remains unknown. 

    \item While both sample show a broad spread in $q$ values, OFGs are skewed towards higher $q$, hinting a preference towards rounded projected morphologies. A near-random distribution on the $A_{v}- q$ plane for both samples may suggest thick disks are observed at the given redshift range of $ 3< z < 4$. The higher $q$ values of OFGs despite similar $n$ to the parent sample could suggest somewhat thicker or dynamically hotter disks--possibly influenced by interactions, minor mergers, or elevated gas fractions, though this remains speculative given current uncertainties. 
    
    \item The mass size relation suggested that OFGs may lie in the intermediate location between SFGs and quiescent galaxies (QGs), although with large scatter. However, further inspecting the compactness using the effective stellar mass surface densities, it is seen that other than higher masses and higher dust attenuation magnitudes in OFGs, there is no significant differences in size or compactness between SFGs and OFGs. While some OFGS may be approaching quiescence as seen by their location on the mass-size relations, most other OFGs show similar characteristics to the parent sample. A comparison of number densities between OFGs and QGs at a lower redshift ($z$) could be used to further investigate if OFGs are indeed in a transition phase between SFGs and QGs or simply an extension of the SFGs towards a high mass end.

\end{enumerate}

\begin{acknowledgements}

This work is based on observations made with the NASA/ESA/CSA James Webb Space Telescope. The raw data were obtained from the Mikulski Archive for Space Telescopes at the Space Telescope Science Institute, which is operated by the Association of Universities for Research in Astronomy, Inc., under NASA contract NAS 5-03127 for \textit{JWST}. 
Some of the data products presented herein were retrieved from the Dawn JWST Archive (DJA). DJA is an initiative of the Cosmic Dawn Center, which is funded by the Danish National Research Foundation under grant No. 140 (DNRF140).
      
This work has received funding from the Swiss State Secretariat for Education, Research and Innovation (SERI) under contract number MB22.00072, as well as from the Swiss National Science Foundation (SNSF) through project grant 200020\_207349. TBM was supported by a CIERA Fellowship.

\end{acknowledgements}

\bibliography{bib} 
\clearpage
\onecolumn

\appendix
\section{Magnitude Cut of 26 mag or 145 nJy}
\label{pysersic_limit}

\begin{figure}[h!]
    \centering
    \includegraphics[width=\linewidth]{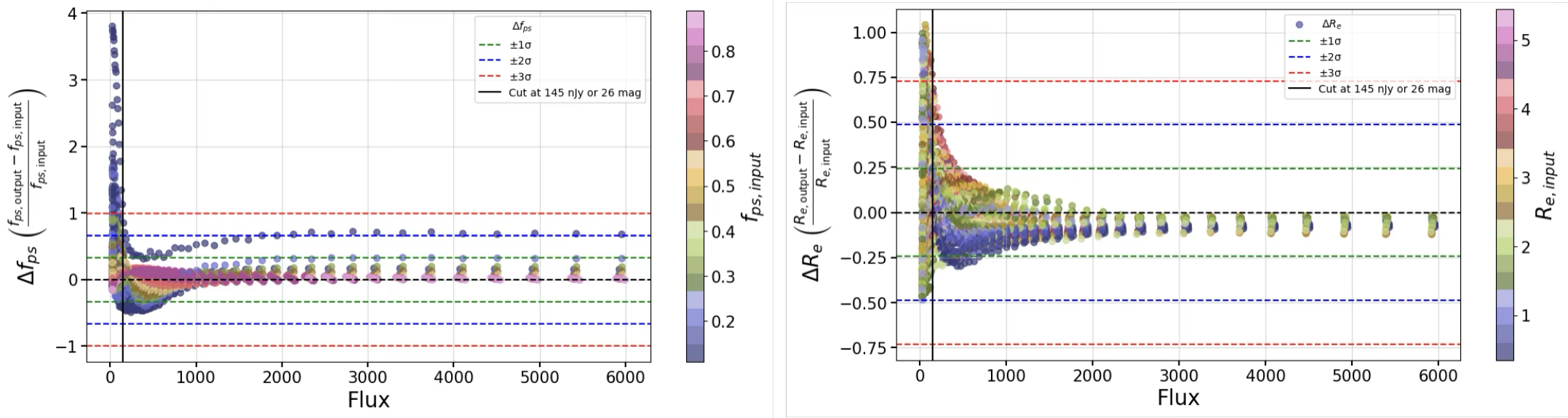}
    \caption{\textit{Left:}Residual plot of PSF fraction $(f_{ps})$ (residual = $\frac{f_{\text{ps, output}}- f_{\text{ps,input}}}{f_{\text{ps,input}}}$
) of all simulated sources measured using \texttt{pysersic} as a function of flux (nJy).The dots are the residuals colour coded by increasing values of $f_{ps,input}$ (\textit{top-left panel}) and Signal-to-noise ratio (SNR) \textit{bottom-left panel}. The black solid vertical line in each plot illustrates the flux limit that has been implemented to select final sample for analysis at 145 nJy or 26 mag.\textit{Right:}Residual plot of effect radius (kpc) (residual= $\frac{R_{\text{e, output}}- R_{\text{e,input}}}{R_{\text{e,input}}}$ ) of Sérsic-dominated ($f_\mathrm{ps}$ <0.5) sources measured with \texttt{pysersic} as a function of flux (nJy).The dots are the residuals colour coded by increasing values of $R_{e, input}$ (\textit{top-right panel}) and SNR \textit{bottom-right panel}.}
     \label{fig:re_res}
\end{figure}

Figure \ref{fig:re_res} justifies the magnitude cuts used for sample selection by assessing the accuracy of measurements with \texttt{pysersic} as a function of flux. To determine the flux limit for reliable size and PSF fraction retrieval, we generated 6000 "fake" sources spanning 60 flux values (22-28 mag), 10 effective radii ($R_e$, 0.5 pix-20 pix), and 10 PSF fractions (0 to 1, fraction of contribution of the PSF to the light profile of the source). The sources were modeled with a Sérsic component(\texttt{astropy.Sersic2D}, $n=1.5, \ q=1$) and a PSF component (PRIMER-UDS field PSF). The "real" Sérsic radius was then calculated using ($(1 - \text{PSF fraction}) \ \times R_e$). To account for the error in fluxes, these models were then injected into a blank sky cutout from the PRIMER-UDS field. Similarly, the RMS (root mean square) map was chosen at the same coordinates as the blank sky map. Additionally, Poisson noise owing to the injected sources was incorporated into the RMS map to create a unique RMS map for each source. This allowed us to account for background noise and statistical fluctuations to accurately assess the uncertainties in the measured flux. The final image, RMS map and PSF had dimensions of 101 $\times$ 101 pixels or 4.04" $\times$ 4.04" for each mock source. The mock images, RMS map and PSF are created the same way in Appendix \ref{gal_vs_py}. We fit a "Sérsic + point source" light profile model to the sources, as we want to recover the PSF fractions and work on "Sérsic-dominated sources" ($< 50\%$ contribution from the PSF to the light profile of the source) for our paper. For the \texttt{pysersic} fitting, we implemented a custom prior setup similar to the one mentioned in the main text in Section \ref{Methods} with and additional free parameter of PSF fraction (0 to 1). Figure \ref{fig:re_res} shows the residual of measurements of PSF fraction ($f_\mathrm{ps}$) and $R_e$ (for $f_\mathrm{ps}<$ 0.5) with respect to the flux of the sources. The residuals are calculated as $\frac{Output - Input}{Input}$ to show error in output values relative to the input values of $R_{e}$ and $f_\mathrm{ps}$. We know that the median sizes for both the parent and OFG samples lie at $\sim 1$ kpc (parent: $1.01 ^{+1.48}_{-1.66}$ kpc, OFG: $0.99 ^{+1.67}_{-1.55}$ kpc). So focusing on this size range, it is evident that for fluxes below 145 nJy (shown by a black solid vertical line in all figures), there is a noticeable increase in the number of outliers with large positive deviations in $\Delta R_e$. For sources around the median sizes observed in the real data, most $R_{e}$ measurements remain around $\pm 2 \sigma$ until flux $>$145 nJy beyond which uncertainties deviate largely. Comparing the $f_\mathrm{ps}$ values, we observe a similar trend where for most values of $f_\mathrm{ps}$, the residuals remain within $2 \sigma$ until 145 nJy and deviate significantly at lower fluxes. Thus, 145 nJy was considered to be the lower limit of flux in the real sources, meaning that sources below this limit were discarded from the sample used for analysis. Note that the deviation in $f_\mathrm{ps}$ residuals for $f_\mathrm{ps} \leqslant 0.1$ exhibits significant outliers. This occurs because we are fitting a Sérsic + point source profile to the sources, which tends to overestimate the point source flux at low $f_\mathrm{ps}$ values. This behaviour is expected from the two-component model fitting priors: at low PSF, the fit overestimates it due to the lack of a substantial PSF component, while at high PSF, the model compensates with a larger Sérsic component. 

\newpage
\section{Kron Radius Criteria}
\label{kron}
\begin{figure}[h!]
    \centering
    \includegraphics[width=1\linewidth]{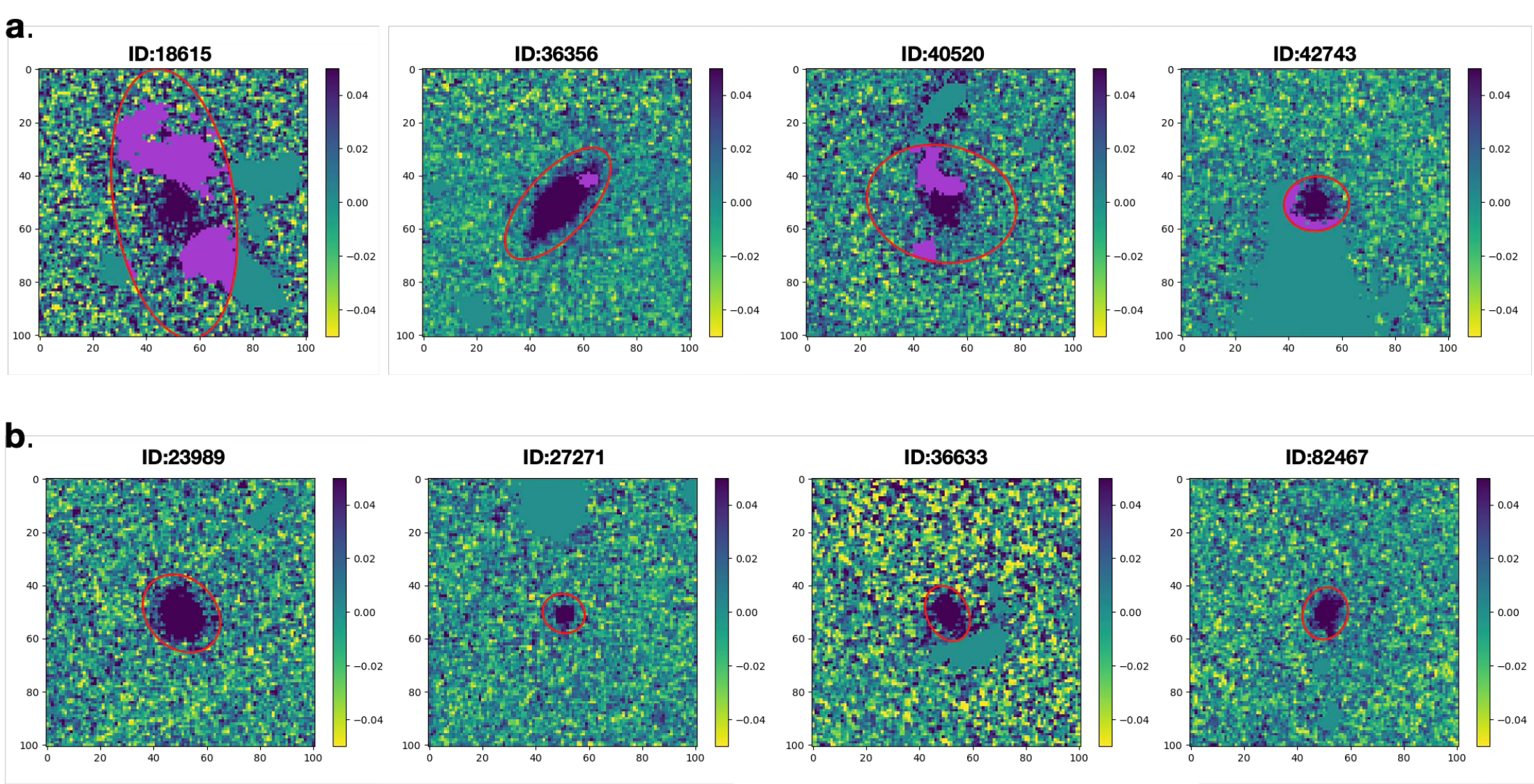}
    \caption{A few example sources to illustrate the difference between "bad" and "good" sources to produce the final sample used for analysis in this paper. Purple coloured pixels in the images denote the zero-valued or masked pixels from neighboring sources.
        \textit{Top panel (a):} Example sources that have been discarded from the sample owing to neighbouring sources within the Kron aperture of the sources, leading to over-masking around the central source. This over-masking can lead to potential inaccuracies in the light profile measurements. \textit{Bottom panel (b)}: Example sources that have been deemed to be perfectly usable for fitting owing to the absence of over-masking or contamination from neighbouring sources within the Kron aperture.}
    \label{fig:Kron}
\end{figure}

Figure \ref{fig:Kron} shows another criteria used to filter out sources, in order to find the best set of sources to perform morphological fitting on. We utilise the Kron radius measurement of the sources as derived from Source-Extractor\citep{SE} to create and aperture boundary around the objects. We check for zero-valued pixels within this boundary, which indicate masked regions used to account for light from neighbouring sources. If the area within this Kron aperture for a given source contains more than three adjacent zero-valued pixels, then this source is discarded from the final sample.
\newpage

\section{\texttt{GALFIT} vs \texttt{pysersic}}
\label{gal_vs_py}
\begin{figure}[h!]
    \centering
    \includegraphics[width=0.8\linewidth]{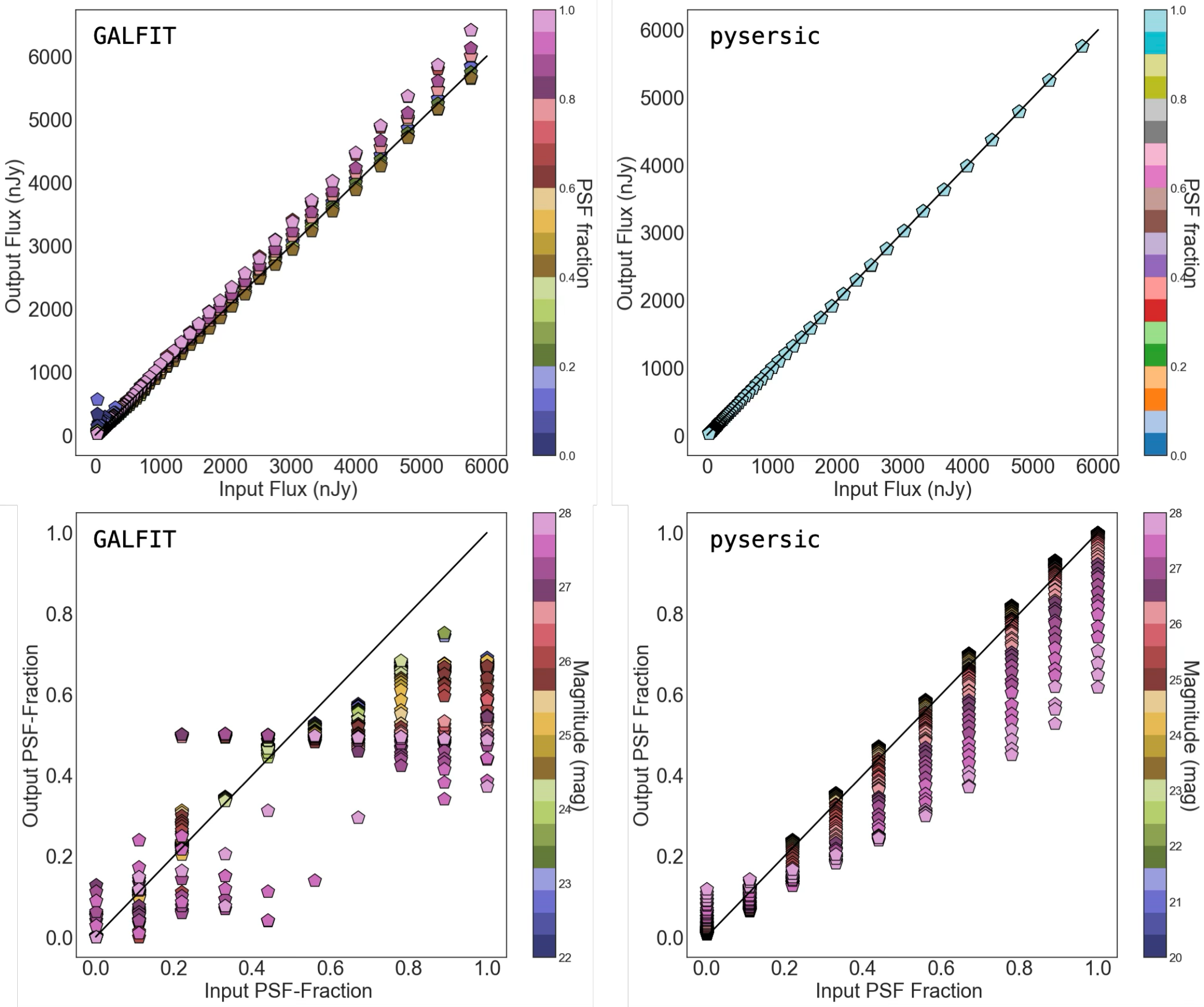}
    \caption{\textit{Top Panel:} Comparison of output flux (y-axis) versus input flux(x-axis) between \texttt{GALFIT} and \texttt{pysersic}. The color bar indicates the PSF-fraction for each source.
    \textit{Bottom Panel:} Comparison of output PSF-fraction (Y-axis) versus input PSF-fraction(X-aixs) between \texttt{GALFIT} and \texttt{pysersic}. The color bar indicates the flux for each source in AB Magnitude.
    }
    \label{fig:gal_pysersic_psfl}
\end{figure}

We evaluate the performance of the two fitting codes \texttt{GALFIT} and \texttt{pysersic} using mock sources of varying flux and PSF fractions (fraction of contribution of the PSF to the light profile of the source). The sources were modeled with a Sérsic component using astropy's \texttt{Sersic2D} functions and a PSF component using the PRIMER-UDS field PSF. A radius of 10 pixels (pixel size: 0.04" $\times$ 0.04") was chosen to create the model for this test. The 10 pixel size was chosen to imply that the sources are well resolved and the only parameter we compare is the quality of reliability of measurement of the sources from the two codes in question. The sources were assigned varying PSF fractions between 0 to 1. The "real" $R_e$ or the radius associated to the Sérsic component was then calculated using ($(1 - \text{PSF fraction}) \ \times 10$). The sources were assumed to have a perfectly face-on orientation $(\text{axis ratio} \ q=1)$ and Sérsic index $(n) = 1.5$. In this way, we generated 600 sources with 60 flux values (fluxes corresponding to magnitudes between 22 and 28 mag) and 10 PSF fractions. The simulated images, RMS map and PSF are created the same way as mentioned in Appendix \ref{pysersic_limit}. The final image, RMS map and PSF had dimensions of 101 $\times$ 101 pixels or 4.04" $\times$ 4.04" for each simulated source. A "Sérsic + point source" light profile model was then fit to all the sources using \texttt{GALFIT} and \texttt{pysersic}.
Both fitting codes were given weakly informative uniform priors. In \texttt{GALFIT}, we allowed the $n$ to range from 0.5 to 5, effective radius from 1.5 to 100 pixels, and $q$ from 0.01 to 1, while fixing the central positions of the Sérsic and PSF components. We know that the PSF fraction represents the contribution of the PSF component to the total flux of the source. So, as a prior, the flux was evenly distributed between the Sérsic and PSF components. In \texttt{pysersic}, input flux values were provided with a small error allowance, and the same parameter constraints as \texttt{GALFIT} were applied, along with the free parameter of PSF fraction (0-1).
Figure \ref{fig:gal_pysersic_psfl} shows the input versus output flux and PSF fraction distribution from both the fitting codes. \texttt{pysersic} more accurately recovers the input distribution, while \texttt{GALFIT} exhibits significant inconsistencies, particularly in PSF fraction retrieval, leading to a less uniform result. This test led us to prefer 
\texttt{pysersic} for subsequent analysis on real sources for the paper.
\newpage
\section{Example Fits}
\label{example}
    \begin{figure}[h!]
        \centering
\includegraphics[width=0.9\linewidth]{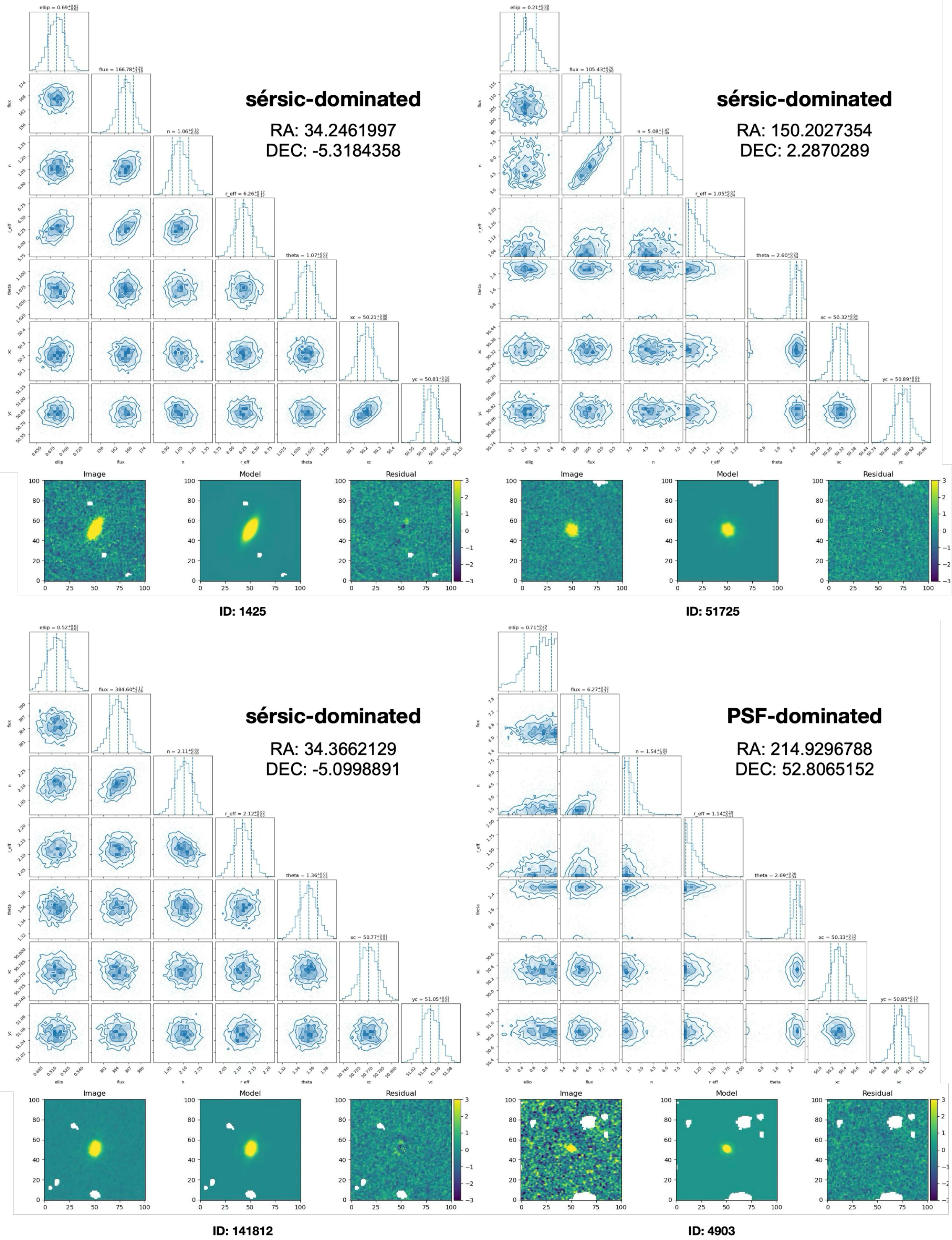}
        \caption{Four examples of Sérsic profile fits of sources with varying morphologies utilising the F444W images from pysersic. The corner plots show the posterior distribution of the inferred parameters. The image panel below each corner plot shows the original image cutout (4" $\times$ 4"), model inferred, and the residual from left to right for each source. The residual is calculated as $\frac{Model - Image}{RMS}$ thus it is a unit-less quantity showing the relative deviations from the noise level.}
        \label{fig:enter-label}
    \end{figure}

\end{document}